\documentstyle[preprint,eqsecnum,aps]{revtex}
\tightenlines
\input epsf

\begin{document}
\draft

\title{CP violating phase and quark mixing angles from flavour permutational symmetry breaking}

\author{A. Mondrag\'on
   and E. Rodr\'{\i}guez-J\'auregui}
\address{
  Instituto de F\'{\i}sica, UNAM, Apdo. Postal 20-364, 01000 M\'exico,
  D.F. M\'exico.}

\date{\today}
\maketitle
\begin{abstract}
The phase equivalence of the theoretical quark mixing matrix ${\bf V}^{th}$ derived from the breaking of the flavour permutational symmetry and the phenomenological parametrizations ${\bf V}^{PDG}$ and ${\bf V}^{KM}$ is explicitly exhibited. From here, we derive exact explicit expressions for the three mixing angles and the CP violating phase of the two phenomenological parametrizations in terms of the quark mass ratios $(m_{u}/m_{t}, m_{c}/m_{t}, m_{d}/m_{b}, m_{s}/m_{b})$ and the parameters $Z^{*1/2}$ and $\Phi^*$ characterizing the preferred symmetry breaking pattern.
The computed values for the CP violating phase and the mixing angles of the standard parametrization advocated by the Particle Data Group are $\delta^*_{13}=73.2^\circ$, $\sin\theta^*_{12}=0.222$, $\sin\theta^*_{13}=0.0036$, and $\sin\theta^*_{23}=0.040$. The computed values of the CP-violating phase and the mixing angles of the Kobayashi-Maskawa parametrization are  $\delta^*_{KM}=96.4^\circ$, $\sin\theta^*_{1}=0.2225$, $\sin\theta^*_{2}=0.0384$, and $\sin\theta^*_{3}=0.0162$. In both cases, the numerical values of the mixing angles and CP-violating phase computed from quark masses and the flavour symmetry breaking parameters coincide almost exactly with the central values of the same parameters obtained from a fit to experimental data.
\end{abstract}


\narrowtext
\section{Introduction}\label{sec:0}
In this paper we review some previous work \cite{ref:1}- \cite{ref:1.1} and give some new results on the functional relations between quark masses and quark mixing angles and CP-violating phases occurring in the phenomenological parametrizations of the quark mixing matrix. These functional relations result from the equivalence under a rephasing of the quark fields of the phenomenological parametrizations and the parametrization derived from the breaking of the flavour permutational symmetry.\\

In the Standard Electroweak theory of particle interactions, quark flavour mixing is described by means of a unitary mixing matrix $\bf V$. The measurables of this matrix, which are invariant under a rephasing of the quark fields, are the moduli of its elements, {\it i. e.}, the quantities $|{\bf V}_{ij}|$. In the case of three families, unitarity of $\bf V$ constrains the number of independent moduli to four. In consequence, phenomenological parametrizations of the quark mixing matrix expressed in terms of four parameters were introduced without taking the possible functional relations between the quark masses and the flavour mixing parameters into account. Kobayashi and Maskawa \cite{ref:1.3} originally chose as independent parameters three rotation angles and one CP-violating phase. A number of parametrizations of this kind, different from the original Kobayashi-Maskawa form have been proposed \cite{ref:1.4}, one of the most commonly used is the ``standard'' parametrization \cite{ref:2} advocated by the Particle Data Group \cite{ref:3}. From a mathematical point of view, two different parametrizations of the $3\times 3$ unitary, quark mixing matrix containing four suitably defined, independent parameters are equivalent if the moduli of corresponding entries are equal. In such scheme, both, the masses of the quarks as well as the mixing angles and the CP-violating phase occur in the theory as free, independent parameters.\\

In contrast, the elements of the quark mixing matrix ${\bf V}^{th}$, derived in two previous papers from a simple ansatz on the breaking of the flavour permutational symmetry \cite{ref:1}-\cite{ref:1.1}, are explicit functions of the four quark mass ratios $m_u/m_t$, $m_c/m_t$, $m_d/m_b$, $m_s/m_b$, and only two, free, linearly independent parameters, namely, the symmetry breaking parameter $Z^{1/2}$ and the CP-violating phase $\Phi$. The numerical values of $Z^{1/2}$ and $\Phi$ which characterize the preferred symmetry breaking pattern are extracted from a $\chi^2$ fit of the theoretical expressions  $|{\bf V}_{ij}^{th}|$ to the experimentally determined values of the moduli of the elements of the mixing matrix  $|{\bf V}_{ij}^{exp}|$. It is remarkable that the quality of the best fit of  ${\bf V}^{th}$ to the experimental data is as good as the quality of the fit of the phenomenological parametrizations ${\bf V}^{PDG}$ or ${\bf V}^{KM}$ to the same data. More precisely, when the best set of parameters of each parametrization is used, the moduli of corresponding entries in ${\bf V}^{th}$ and ${\bf V}^{PDG}$ or ${\bf V}^{KM}$  are numerically equal and give an equally good representation of the experimentally determined values of the moduli of the mixing matrix $|{\bf V}_{ij}^{exp}|$. Hence, we are justified in writing 

\begin{eqnarray}\label{eq:1}
|{\bf V}^{th}_{ij}|= |{\bf V}^{PDG}_{ij}|\quad or\quad |{\bf V}^{th}_{ij}|= |{\bf V}^{KM}_{ij}|,  
\end{eqnarray}
even though ${\bf V}^{th}$ has only two free, real linearly independent adjustable parameters, while the number of adjustable parameters in ${\bf V}^{PDG}$ or  ${\bf V}^{KM}$ is four.\\

In what follows, it will be shown that by means of suitable rephasing of the quark fields ${\bf V}^{th}$  may be changed into new, phase transformed forms $\tilde{\bf V}^{th}$ or $\hat{\bf V}^{th}$ such that all their matrix elements are numerically equal to  the corresponding entries in ${\bf V}^{PDG}$ or ${\bf V}^{KM}$, both in modulus and phase. Once this equality is established, we derive exact explicit analitycal expressions for the mixing angles and the CP-violating phase of the two phenomenological parametrizations ${\bf V}^{PDG}$ and ${\bf V}^{KM}$, in terms of the quark mass ratios $m_u/m_t$, $m_c/m_t$, $m_d/m_b$, $m_s/m_b$, the flavour symmetry breaking parameter $Z^{1/2}$ and the CP-violating phase $\Phi$.\\
 The plan of this paper is as follows: In Sec. \ref{sec:2}, we introduce some basic concepts and fix the notation by way of a very brief sketch of the group theoretical derivation of mass matrices with a modified Fritzsch texture. Sect.~\ref{sec:3} is devoted to the derivation of exact, explicit expressions for the elements of the mixing matrix $V^{th}_{ij}$ in terms of the quark mass ratios and the parameters $Z^{1/2}$ and  $\Phi$ characterizing the symmetry breaking pattern. In Sec.~\ref{sec:4.0}, the phase equivalence of ${\bf V}^{th}$ and the phenomenological parametrrization ${\bf V}^{PDG}$ and ${\bf V}^{KM}$ is established. The equations that define the rephasing transformation connecting ${\bf V}^{th}$ and ${\bf V}^{PDG}$ are solved in Sec.~\ref{sec:4}. Sections ~\ref{sec:5} and ~\ref{sec:6} are devoted to the derivation of explicit expressions for the mixing parameters $\sin\theta_{12}$, $\sin\theta_{23}$, $\sin\theta_{13}$ and the CP violating phase $\delta_{13}$ of ${\bf V}^{PDG}$ as funtions of the quark mass ratios and the parameters $Z^{1/2}$ and $\Phi$. In Sec.~\ref{sec:7}, the equations that define the rephasing transformation relating ${\bf V}^{th}$ and ${\bf V}^{KM}$ are solved. Explicit expressions for the mixing parameters $\sin\theta_{1}$, $\sin\theta_{2}$, $\sin\theta_{3}$ and the CP violating phase $\delta_{KM}$ of ${\bf V}^{KM}$ as funtions of the quark mass ratios and the symmetry breaking parameters are obtained in Sec.~\ref{sec:7.1}  and ~\ref{sec:7.2}. Our paper ends in Sec.~\ref{sec:11} with a summary of results and some conclusions.   

\section{Mass matrices from the breaking of $S_{L}(3)\otimes S_{R}(3)$ }
\label{sec:2}

In the Standard Model, analogous fermions in different generations,
say ${\it u,c}$ and ${\it t}$ or ${\it d,s}$ and ${\it b}$, have
completely identical couplings to all gauge bosons of the strong, weak
and electromagnetic interactions.  Prior to the introduction of the
Higgs boson and mass terms, the Lagrangian is chiral and invariant
with respect to any permutation of the left and right quark fields.
The introduction of a Higgs boson and the Yukawa couplings give mass
to the quarks and leptons when the gauge symmetry is spontaneously
broken. The quark mass term in the Lagrangian, obtained by taking the
vacuum expectation value of the Higgs field in the quark Higgs
coupling, gives rise to quark mass matrices ${\bf M_d}$ and ${\bf
  M_u}$,

\begin{equation}\label{eq:21}
{\cal L}_{Y} ={\bf \bar{q}}_{d,L}{\bf M}_{d}{\bf q}_{d,R}+
{\bf\bar{q}}_{u,L}{\bf M}_{u}{\bf q}_{u,R}+h.c.
\end{equation}

In this expression, ${\bf q}_{d,L,R}(x)$ and ${\bf q}_{u,L,R}(x)$ denote the
left and right quark $d$- and $u$-fields in the current or weak 
basis, ${\bf q}_{q}(x)$ is a column matrix, its components ${\bf q}_{q,k}(x)$ are the quark Dirac fields, $k$ is the flavour index. In this basis, the charged hadronic currents,

\begin{equation}
\label{eq:23}
J_{\mu}\sim \bar{q}_{u,L}\gamma _{\mu}q_{d,L},
\end{equation}
are not changed if both, the d-type and the u-type fields are transformed with the same unitary matrix.

\subsection{Modified Fritzsch texture}
\label{sec:2.1}
A number of authors \cite{ref:1}-\cite{ref:1.1}, [\cite{ref:5}-\cite{ref:23}] have pointed out that realistic quark mass matrices result from the flavour permutational symmetry $S_{L}(3)\otimes S_{R}(3)$ and its spontaneous or explicit breaking.
The group $S(3)$ treats three objects symmetrically, while the
hierarchical nature of the mass matrices is a consequence of the
representation structure $\bf{1\oplus2}$ of $S(3)$, which treats the
generations differently.
Under exact $S_{L}(3)\otimes S_{R}(3)$ symmetry, the mass spectrum for either up or down quark sectors consists of one massive particle in a singlet irreducible representation and a pair of massless particles in a doublet irreducible representation, the corresponding quark mass matrix with the exact $S_{L}(3)\otimes S_{R}(3)$ symmetry will be denoted by ${\bf M}_{3q}$. In order to generate masses for the first and second families, we add the terms ${\bf M}_{2q}$ and ${\bf M}_{1q}$ to ${\bf M}_{3q}$. The term ${\bf M}_{2q}$ breaks the permutational symmetry  $S_{L}(3)\otimes S_{R}(3)$ down to $S_{L}(2)\otimes S_{R}(2)$ and mixes the singlet and doublet representation of $S(3)$. ${\bf M}_{1q}$ transforms as the mixed symmetry term in the doublet complex tensorial representation of $S_{diag}(3) \subset S_{L}(3)\otimes S_{R}(3)$. Putting the first family in a complex representation will allow us to have a CP violating phase in the mixing matrix. Then, in a symmetry adapted basis , ${\bf M}_{q}$ takes the form

\begin{equation}\label{eq:2.11}
\begin{array}{rcl}
{M_{q}}&=&{m_{3q}}\left[ \pmatrix{
0 & {A_{q}}e^{-i\phi_{q}} & 0 \cr
{A_{q}}e^{i\phi_{q}} & 0 & 0 \cr
0 & 0 & 0 \cr
}+\pmatrix{
0 & 0 & 0 \cr
0 & -\triangle_{q}+\delta_{q} & B_{q} \cr
0 & B_{q} & \triangle_{q}-\delta_{q} \cr
}\right]\\ &+&m_{3q}\pmatrix{
0 & 0 & 0 \cr
0 & 0 & 0 \cr
0 & 0 & 1-\triangle_{q} \cr
}=m_{3q}\pmatrix{
0 & A_{q}e^{-i\phi_{q}} & 0 \cr
A_{q}e^{i\phi_{q}} & -\triangle_{q}+\delta_{q} & B_{q} \cr
0 & B_{q} & 1-\delta_{q} \cr
}.
\end{array}
\end{equation}

From the strong hierarchy in the masses of the quark families, 
$m_{3q}>> m_{2q}> m_{1q}$, we expect $1-\delta_{q}$ to be very close to unity.
The entries in the mass matrix may be readily expressed in terms of the mass eigenvalues $(m_{1q}, -m_{2q}, m_{3q})$ and the small parameter $\delta_{q}$.
 Computing the invariants of $M_{q}$, $tr M_{q}$, $tr {M_{q}}^{2}$ and $det M_{q}$, we get
\begin{eqnarray}\label{eq:2.12}
A^{2}_{q}={\tilde m_{1q}}{\tilde m_{2q}}(1-\delta _{q})^{-1}\qquad ,\qquad
\triangle _{q}= {\tilde m_{2q}}-{\tilde m_{1q}},\\ \cr
B^{2}_{q}=\delta_{q}((1-\tilde m_{1q}+\tilde m_{2q}-\delta _{q})-
\tilde m_{1q}{\tilde m_{2q}}(1-\delta _{q})^{-1}),
\end{eqnarray}
where 
${\tilde m_{1q}}={m_{1q}/m_{3q}}$ and 
${\tilde m_{2q}}={m_{2q}/m_{3q}}$.

If each possible symmetry breaking pattern is now characterized by the ratio
\begin{equation}\label{eq:2.13}
{Z_{q}}^{1/2}={B_{q}/(-\triangle _{q}+\delta_{q})},
\end{equation}
the small parameter $\delta _{q}$ is obtained as the solution of the cubic equation

\begin{equation}\label{eq:2.14}
\delta_{q}\left[ (1+\tilde m_{2q}- \tilde m_{1q}- \delta_{q})(1-\delta_{q})-
{\tilde m_{1q}}{\tilde m_{2q}}\right] - Z_{q}(-{\tilde m_{2q}}+{\tilde m_{1q}}+
\delta_{q})^{2}=0,
\end{equation}
which vanishes when $Z_{q}$ vanishes. An exact explicit expression for $\delta_q$ as function of the quark mass ratios and $Z_q$ is given in \cite{ref:1}. 
 An approximate solution to Eq. (\ref{eq:2.14}) for $\delta_{q}(Z_{q})$, valid for small
values of $Z_{q}$ ($Z_{q}\leq 10$), is
\begin{eqnarray}\label{eq:2.15}
 \delta_{q}\left( Z_{q} \right)\approx {Z_{q}
\left(   \tilde{m}_{2q}-\tilde{m}_{1q} \right)^{2}
\over \left(1-\tilde{m}_{1q} \right)\left( 1  +
\tilde{m}_{2q} \right)+2Z_{q}\left(   \tilde{m}_{2q}-\tilde{m}_{1q} 
\right)(1+{1\over 2}(\tilde{m}_{2q}-\tilde{m}_{1q}))}~.
\end{eqnarray}
\subsection{Symmetry breaking pattern}
\label{sec:2.2}
In the symmetry adapted basis, the matrix $\bf{M}_{2q}$, written in term of $Z^{1/2}_q$, takes the form

\begin{equation}\label{eq:2.21}
{\bf M}_{2q} =
m_{3q}\left(-\tilde{m}_{2q}+\tilde{m}_{1q}+\delta_{q}\right)\pmatrix{
0 & 0 & 0 \cr
0 & 1 & Z^{1/2}_q \cr
0 & Z^{1/2}_q & -1 \cr
},
\end{equation}
when $Z^{1/2}_q$ vanishes, ${\bf M}_{2q}$ is diagonal and there is no mixing of singlet and doublet representations of $S(3)$. Therefore, in the symmetry adapted basis, the parameter $Z^{1/2}_q$ is a measure of the amount of mixing of singlet and doublet irreducible representations of $S_{diag}(3)\subset S_L(3) \otimes S_R(3)$.\\

We may easily give a meaning to $Z^{1/2}_q$ in terms of permutations. From Eqs. (\ref{eq:21}) and (\ref{eq:2.21}), we notice that the symmetry breaking term in the Yukawa Lagrangian, $\bar{\bf q}_L {\bf M}_{2q}{\bf q}_R $ is a functional of only two fields:
$\frac{1}{\sqrt 3}\left( q_2(X)+\sqrt{2} q_3(X)\right)$ and $\frac{1}{\sqrt 3}\left(-\sqrt{2} q_2(X)+ q_3(X)\right)$.
Under the permutation of these fields, $\bar{\bf q}_L {\bf M}_{2q}{\bf q}_R $ splits into the sum of an antisymmetric term $\bar{\bf q}_L {\bf M}^A_{2q}{\bf q}_R $ which changes sign, and a symmetric term $\bar{\bf q}_L {\bf M}^S_{2q}{\bf q}_R $, which remains invariant,
\begin{equation}\label{eq:2.23}
{\bf M}_{2q}=-\frac{2}{9}m_{3q}\bigg\{a\pmatrix{
0 & 0 & 0 \cr
0 & 1 & -\sqrt 8 \cr
0 & -\sqrt 8 & -1 \cr
}+ 2b\pmatrix{
0 & 0 & 0 \cr
0 & 1 & \frac{1}{\sqrt 8} \cr
0 & \frac{1}{\sqrt 8} & -1 \cr
}\bigg\},
\end{equation}
where $a=(\delta_q-\triangle_q)(\sqrt 2 Z^{1/2}_q -\frac{1}{2})$ and $b=(\delta_q-\triangle_q)(\frac{\sqrt 2}{2} Z^{1/2}_q +{2})$.
It is evident that there is a corresponding decomposition of the mixing parameter $Z^{1/2}_q$,
\begin{equation}\label{eq:2.25}
{Z_{q}}^{1/2}=N_{Aq}Z^{1/2}_A+N_{Sq}Z^{1/2}_S
\end{equation}
 with 

\begin{equation}\label{eq:2.27}
1=N_{Aq}+N_{Sq},
\end{equation}
where $Z^{1/2}_A=-\sqrt 8$ is the mixing parameter of the matrix ${\bf M}^A_{2q}$, and $Z^{1/2}_S=\frac{1}{\sqrt 8}$ is the mixing parameter of ${\bf M}^S_{2q}$.
In this way, a unique linear combination of $Z^{1/2}_{A}$ and $Z^{1/2}_{S}$  is associated to the simmetry breaking pattern characterized by $Z^{1/2}_{q}$. Thus, the different symmetry breaking patterns defined by ${\bf M}_{2q}$ for different values of the mixing parameter $Z^{1/2}_{q}$ are labeled in terms of the irreducible representations of the group $\tilde S(2)$ of permutations of the two fields in $\bar{\bf q}_L {\bf M}_{2q}{\bf q}_R $. 
The pair of numbers $(N_A, N_S)$ enters as a convenient mathematical label of the symmetry breaking pattern without introducing any assumption about the actual pattern of $S_L(3)\otimes S_R(3)$ symmetry breaking realized in nature.
\subsection{The Jarlskog invariant}
\label{sec:2.3}

The Jarlskog invariant, $J$, may be computed directly from the commutator of the mass matrices ~\cite{ref:4}

\begin{equation}\label{eq:2.31}
J=- \frac{det \{ -i[{\bf M}_{u}, {\bf M}_{d}]\} }{2F}
\end{equation}
where 
\begin{equation}\label{eq:2.33}
F=(1+\tilde{m}_{c})(1-\tilde{m}_{u})(\tilde{m}_{c}+\tilde{m}_{u})
(1+\tilde{m}_{s})(1-\tilde{m}_{d})(\tilde{m}_{s}+\tilde{m}_{d}).
\end{equation}
Substitution of the expression (\ref{eq:2.11}) for ${\bf M}_u$ and ${\bf M}_d$, in Eq. (\ref{eq:2.31}), with $Z^{1/2}_u=Z^{1/2}_d=Z^{1/2}$ gives

\begin{eqnarray}\label{eq:2.35}
&J&={{Z  \sqrt{{\tilde m_{u}/ \tilde m_{c}}
\over{1-\delta_{u}}} \sqrt{{\tilde m_{d}/ \tilde m_{s}}
\over {1-\delta_{d}}}sin{\Phi}}\over{(1+\tilde m_{c})
(1-\tilde m_{u})(1+\tilde m_{u}/ \tilde m_{c})
(1+\tilde m_{s})(1-\tilde m_{d})(1+\tilde m_{d}/ \tilde m_{s})}}\cr
& \times &\bigg\{
 [(-\triangle_{u}+\delta_{u})(1-\delta_{d})-(-\triangle_{d}+\delta_{d})(1-\delta_{u})]^{2}-{\left({\tilde m_{u}\tilde m_{c}}\over {1-\delta_{u}}\right)}(-\triangle_{d}+\delta_{d})^{2}\cr & - & {\left({\tilde m_{d}\tilde m_{s}\over {1-\delta_{d}}}\right)}(-\triangle_{u}+\delta_{u})^{2}+2 \sqrt{\tilde m_{u}\tilde m_{c}\over {1-\delta_{u}}}\sqrt{\tilde m_{d}\tilde m_{s}\over {1-\delta_{d}}}(-\triangle_{u}+\delta_{u})(-\triangle_{d}+\delta_{d})cos{\Phi}\bigg\}.
\end{eqnarray}
where $\triangle _q$ and $\delta_q$ are defined in Eqs. (\ref{eq:2.12}) and (\ref{eq:2.14}).
In this, way, an exact closed expression for $J$ in terms of the quark mass ratios, the CP violating phase $\Phi$, and the parameter $Z$ that characterizes the symmetry breaking pattern is derived.

\section{The Mixing Matrix}\label{sec:3}
The Hermitian mass matrix ${\bf M}_{q}$ may be written in terms of a real symmetric matrix ${\bf\bar M}_{q}$ and a diagonal matrix of phases ${\bf P}_{q}$ as follows

\begin{equation}\label{eq:31}
{\bf M}_{q}={\bf P}_{q}{{\bf\bar M}_{q}}{{\bf P}_{q}}^{\dagger},
\end{equation}
The real symmetric matrix ${\bf\bar M}_{q}$ may be brought to a diagonal form by means of an orthogonal transformation 
\begin{equation}\label{eq:33}
{\bf\bar M}_{q}={\bf O}_{q}{\bf M}_{q, diag}{\bf O}^{T}_{q},
\end{equation}
where
\begin{equation}\label{eq:35}
{\bf M}_{q, diag}=m_{3q}~diag\left[ ~\tilde m_{1q},~-\tilde m_{2q}, ~1\right],
\end{equation}
with subscripts $1, 2, 3$ refering to $u, c, t$ in the u-type sector and $d, s , b$ in the d-type sector.\\
After diagonalization of the mass matrices ${\bf M}_q$, one obtains the mixing matrix ${\bf V}^{th}$ as

\begin{equation}\label{eq:37}
{\bf V}^{th}={{\bf O}_{u}}^{T}{\bf P}^{u-d}{\bf O}_{d}, 
\end{equation}
where ${\bf P}^{u-d}$ is the diagonal matrix of relative phases

\begin{equation}\label{eq:39}
{\bf P}^{u-d}=diag[1,e^{i\Phi},e^{i\Phi}],
\end{equation} 
and 
\begin{equation}\label{eq:311}
\Phi=(\phi_{u}-\phi_{d}).
\end{equation} 
The orthogonal matrix ${\bf O}_{q}$is given by

\begin{equation}\label{eq:313}
\begin{array}{rcl}
{\bf O}_{q}= \pmatrix{
({\tilde m_{2q}}f_{1}/D_{1})^{1/2} & - ({\tilde m_{1q}}f_{2}/D_{2})^{1/2} & 
({\tilde m_{1q}}{\tilde m_{2q}}f_{3}/D_{3})^{1/2} \cr 
((1-\delta_{q}){\tilde m_{1q}}f_{1}/D_{1})^{1/2} & ((1-\delta_{q}){\tilde m_{2q}}f_{2}/D_{2})^{1/2} & ( (1-\delta_{q}) f_{3}/D_{3})^{1/2} \cr
-({\tilde m_{1q}}f_{2}f_{3}/D_{1})^{1/2} & 
-({\tilde m_{2q}}f_{1}f_{3}/D_{2})^{1/2} & ( f_{1}f_{2}/D_{3})^{1/2} \cr 
},
\end{array}
\end{equation}
where
\begin{eqnarray}\label{eq:315}
{\rm {f}}_{1}=1-\tilde{m}_{1q}-{\delta}_{q},\quad\quad
{\rm {f}}_{2}=1+\tilde{m}_{2q}-{\delta}_{q},\quad\quad
{\rm {f}}_{3}={\delta}_{q},
\end{eqnarray}

\begin{eqnarray}\label{eq:317}
{D}_{1}=(1-\delta_q)\left( 1-\tilde{m}_{1q} \right)\left( 
\tilde{m}_{2q}+\tilde{m}_{1q} \right),
\end{eqnarray}

\begin{eqnarray}\label{eq:319}
{D}_{2}=(1-\delta_q)\left( 1+\tilde{m}_{2q} \right)\left( 
\tilde{m}_{2q}+\tilde{m}_{1q} \right),
\end{eqnarray}

\begin{eqnarray}\label{eq:321}
{D}_{3}=(1-\delta_q)\left( 1+\tilde{m}_{2q} \right)\left( 
1-\tilde{m}_{1q} \right).
\end{eqnarray}

In these expressions, $\delta_u$ and $\delta_d$ are, in principle, functions of the quark mass ratios and the parameters $Z^{1/2}_u$ and $Z^{1/2}_d$ respectively. However, in \cite{ref:1} we found that keeping  $Z^{1/2}_u$ and $Z^{1/2}_d$ as free, independent parameters gives rise to a continuous ambiguity in the fitting of $|V^{th}_{ij}|$ to the experimental data. To avoid this ambiguity we further assumed that the up and down mass matrices are generated following the same symmetry breaking pattern, that is,
\begin{equation}\label{eq:323}
Z^{1/2}_u=Z^{1/2}_d=Z^{1/2}.
\end{equation}
Then, from Eqs. (\ref{eq:37}) - (\ref{eq:323}) all matrix elements in ${\bf V}^{th}$ may be written in terms of four quark mass ratios and only two free, real parameters: the parameter $Z^{1/2}$ which characterizes the symmetry breaking pattern in the u- and d-sectors and the CP violating phase $\Phi$.
The computation of $V^{th}_{ij}$ is quite straightforward. Here, we will give, in explicit form, only those elements of  ${\bf V}^{th}$ which will be of use later. From Eqs. (\ref{eq:37})-(\ref{eq:323}) we obtain, 
\begin{eqnarray}\label{eq:325}\hspace{1.5cm}
V^{th}_{us}&=&-\left(\frac{ \tilde {m}_c \left(1-\tilde{m}_u -\delta_u \right)
\tilde {m}_d\left(1+\tilde{m}_s -\delta_d \right) }
{\left( 1-\delta_u \right) \left( 1-\tilde{m}_u \right)
\left( \tilde{m}_c +\tilde{m}_u \right)\left(1-\delta_d \right)
\left( 1+\tilde{m}_s\right)
\left( \tilde{m}_s + \tilde{m}_d\right)}\right)^{1/2} \cr &+&
\left( \frac{\tilde{m}_u \tilde{m}_s}
{\left( 1-\tilde{m}_u \right)\left( \tilde{m}_c+
\tilde{m}_u \right)\left(\tilde{m}_d +\tilde{m}_s
\right)} \right) ^{1/2}
\bigg\{ \left(\frac {\left(1-\tilde{m}_u - \delta_u \right)
\left(1+\tilde{m}_s- \delta_d \right)}{\left( 1+\tilde{m}_s \right)}
\right)^{1/2}\cr &+&
\left(\frac{ \left(1+\tilde{m}_c- \delta_u \right)\delta_u
\left(1-\tilde{m}_d - \delta_d \right)\delta_d }
{(1-\delta_u)(1-\delta_d)(1+\tilde{m}_s)}\right)^{1/2}
\bigg\}e^{i\Phi}
\end{eqnarray}

\begin{eqnarray}\label{eq:327}
V^{th}_{ub}&=&\left( \frac{\tilde {m}_c (1-\tilde{m}_u -\delta_u ) }{ (1-\delta_u)
(1-\tilde{m}_u)(\tilde{m}_c + \tilde{m}_u)} \frac {\tilde {m}_d \tilde {m}_s
\delta_d }{(1- \delta_d )(1+ \tilde {m}_s)(1 - \tilde {m}_d) } \right)^{1/2}
\cr
&+&\bigg\{- \left(\frac{\tilde{m}_u (1+\tilde{m}_c -\delta_u)\delta_u 
(1-\tilde{m}_d -\delta_d)(1+\tilde{m}_s -\delta_d)}
{(1- \delta_u )(1-\tilde{m}_u)(\tilde{m}_c +\tilde{m}_u)(1- \delta_d)
(1+\tilde{m}_s)(1-\tilde{m}_d)}
\right)^{1/2}
\cr &+& \left(\frac{\tilde{m}_u (1-\tilde{m}_u -\delta_u)\delta_d}
{(1-\tilde{m}_u)(\tilde{m}_c +\tilde{m}_u)(1+\tilde{m}_s)(1-\tilde{m}_d)}
\right)^{1/2}\bigg\}e^{i\Phi}  
\end{eqnarray}

\begin{eqnarray}\label{eq:329}
V^{th}_{cs}&=&\left(\frac{\tilde {m}_u \left( 1+\tilde{m}_c -\delta_u \right)
\tilde {m}_d \left( 1+\tilde{m}_s -\delta_d \right)}{\left( 1-\delta_u \right)
\left( 1+\tilde{m}_c \right)\left( \tilde {m}_c+\tilde {m}_u \right)
\left( 1-\delta_d \right)\left( 1+\tilde{m}_s \right)
\left( \tilde {m}_s+\tilde {m}_d \right)}\right)^{1/2}\cr &+& \bigg\{
\left(\frac{\tilde {m}_c\delta_u\left( 1-\tilde {m}_u-\delta_u \right)
\tilde {m}_s\delta_d
\left( 1-\tilde{m}_d -\delta_d \right)}{\left( 1-\delta_u \right)
\left( 1+\tilde{m}_c \right)\left( \tilde {m}_c+\tilde {m}_u \right)
\left( 1-\delta_d \right)\left( 1+\tilde{m}_s \right)
\left( \tilde {m}_s+\tilde {m}_d \right)}\right)^{1/2}\cr &+&
\left(\frac{\tilde {m}_c
\left( 1+\tilde{m}_c -\delta_u \right)\tilde {m}_s
\left( 1+\tilde{m}_s -\delta_d \right)}{
\left( 1+\tilde{m}_c \right)\left(1-\tilde {m}_u \right)
\left( 1+\tilde{m}_s \right)
\left( 1-\tilde {m}_d \right)}\right)^{1/2}
\bigg\}e^{i\Phi}
\end{eqnarray}

and 

\begin{eqnarray}\label{eq:331}
{V^{th}_{cb}}&=&- \left( \frac{\tilde {m}_u (1+\tilde{m}_c -\delta_u ) }
{ (1-\delta_u)(1+\tilde{m}_c)(\tilde{m}_c + \tilde{m}_u)} \frac {
\tilde {m}_d \tilde {m}_s \delta_d }{(1- \delta_d )(1+ \tilde {m}_s)
(1 - \tilde {m}_d) } \right)^{1/2}\cr
&+& \bigg\{- \left(\frac{\tilde{m}_c (1-\tilde{m}_u -\delta_u)\delta_u 
(1-\tilde{m}_d -\delta_d)(1+\tilde{m}_s -\delta_d)}
{(1- \delta_u )(1+\tilde{m}_c)(\tilde{m}_c +\tilde{m}_u)(1- \delta_d)
(1+\tilde{m}_s)(1-\tilde{m}_d)}
\right)^{1/2} 
\cr &+&\left(\frac{\tilde{m}_c (1+\tilde{m}_c -\delta_u)}
{(\tilde{m}_c +\tilde{m}_u)(1+\tilde{m}_c)}\frac{\delta_d}{(1+\tilde{m}_s)
(1-\tilde{m}_d)}\right)^{1/2}\bigg\}e^{i\Phi}.
\end{eqnarray}
\subsection{The ``best'' symmetry breaking pattern}
\label{sec:3.1}
In order to find the actual pattern of $S_L(3)\otimes S_R(3)$ symmetry breaking realized in nature,we made a $\chi^2$ fit of the exact expressions for the moduli of the entries in the mixing matrix, $|V^{th}_{ij}|$, the Jarlskog invariant, $J^{th}$, and the three inner angles of the unitarity triangle, $\alpha^{th}$, $\beta^{th}$ and $\gamma^{th}$, to the experimentally determined values of $|{ V}^{exp}_{ij}|$, $J^{exp}$, $\alpha^{exp}$, $\beta^{exp}$ and $\gamma^{exp}$. A detailed account of the fitting procedure is given in \cite{ref:1}. Here, we will give only a brief relation of the main points in the fitting procedure.\\
For the purpose of calculating quark mass ratios and computing the mixing matrix, it is convenient to give all quark masses as running masses at some common energy scale \cite{ref:24}, \cite{ref:25}. In the present calculation, following Peccei \cite{ref:24}, Fritzsch \cite{ref:26} and the Ba-Bar book \cite{ref:27}, we used the values of the running quark masses evaluated at $\mu=m_t$.

\begin{eqnarray}\label{eq:3.10}
m_u&=&3.25\pm 0.9~MeV\quad\quad 
m_c=760\pm 29.5~MeV\quad\quad 
m_t=171.0\pm 12~GeV
\cr
m_d&=&4.4\pm 0.64~MeV\quad\quad~~
m_s=100\pm 6~MeV\quad\quad~~ 
m_b=2.92\pm 0.11~GeV
\end{eqnarray}
These values, with the exception of $m_s$, $m_c$ and $m_b$, were taken from the work of Fusaoka and Koide \cite{ref:25} see also Fritzsch \cite{ref:26} and Leutwyler \cite{ref:28}. The values of $m_c(m_t)$ and $m_b(m_t)$ were obtained by rescaling to $\mu=m_t$ the recent calculations of $m_c(m_c)$ and $m_b(m_b)$ by Pineda and Yndur\'ain \cite{ref:29} and Yndur\'ain \cite{ref:30}. The value of $m_s$
 agrees with the latest determination made by the ALEPH collaboration from a study of $\tau$ decays involving kaons \cite{ref:31}. \\
We kept the mass ratios $\tilde m_c=m_c/m_t$ , $\tilde m_s=m_s/m_b$ and $\tilde m_d=m_d/m_b$ fixed at their central values

\begin{eqnarray}\label{eq:3.11}
\tilde m_c=0.0044, \quad\quad\tilde m_s=0.034 \quad\quad ~and 
\quad\quad\tilde m_d=0.0015,
\end{eqnarray}
but we took the value 

\begin{eqnarray}\label{eq:3.13}
\tilde m_u=0.000036,
\end{eqnarray}
which is close to its upper bound.
We found the following best values for $\Phi$ and $Z^{1/2}$,

\begin{eqnarray}
\Phi^{*}=90^{\circ},\quad\quad Z^{*1/2}=\frac{1}{2}\left[Z^{1/2}_S-Z^{1/2}_A\right]=\sqrt\frac{81}{32}.
\label{eq:3.15}
\end{eqnarray}
corresponding to a value of $\chi^2\leq 0.32$.  The values of the parameters $\delta_u(Z)$ and $\delta_d(Z)$ obtained from (\ref{eq:3.11}), (\ref{eq:3.13}) and  (\ref{eq:3.15}) are
\begin{eqnarray}\label{eq:3.16}
\delta_u(Z^{*1/2})=0.000048, \quad\quad\quad \delta_d(Z^{*1/2})=0.00228.
\end{eqnarray}
Before proceeding to give the numerical results for the mixing matrix ${\bf V}^{th}$, it will be convenient to stress the following points:
\begin{enumerate}
\item
The masses of the lighter quarks are the less well determined, while the moduli of the entries in $|V^{exp}_{ij}|$ with the largest error bars, namely $|V_{ub}|$ and $|V_{td}|$, are the most sensitive to changes in the ratios $m_u/m_c$ and $m_d/m_s$ respectively. Hence, the quality of the fit of $|V^{th}_{ij}|$ to $|V^{exp}_{ij}|$ is good $(\chi^2\leq 0.5)$ even if relatively large changes in the masses of the lighter quarks are made. The sensitivity of $|V_{ub}|$ and $|V_{td}|$ to changes in  $m_u/m_c$ and $m_d/m_s$ respectively, is reflected in the shape of the unitarity triangle which changes appreciably when the masses of the ligther quarks change within their uncertainty ranges. The best simultaneous $\chi^2$ fit of $|V^{th}_{ij}|$, $J^{th}$, and $\alpha^{th}$, $\beta^{th}$ and $\gamma^{th}$, to the experimentally determined quantities was obtained when the ratio  $\tilde m_u=m_u/m_t$ is taken close to its upper bound, as given in (\ref{eq:3.13}). Furthermore, the chosen high value of $\tilde m_u$ gives for the ratio $|V_{ub}/V_{cb}|$ the value 

\begin{eqnarray}\label{eq:3.17}
\frac{|V_{ub}|}{|V_{cb}|}\approx \sqrt{\frac{m_u}{m_c}}=0.09\pm 0.025
\end{eqnarray}
in very good agreement with its latest world average [\cite{ref:32}, \cite{ref:33}, \cite{ref:34}].
\item
As the energy scale changes, say from $\mu=m_t$ to $\mu=1~GeV$, the running quark masses change appreciably, but since the masses of light and heavy quarks increase almost in the same proportion, the resulting dependence of the quark mass ratios on the energy scale is very weak. When the energy scale changes from  $\mu=m_t$ to $\mu=1~GeV$, $\tilde m_u$ and $\tilde m_d$ decrease by about $25\%$ and $\tilde m_c$ and $\tilde m_s$ also decrease but by less than $16\%$.
\item
In view of the previous considerations, a reasonable range of values for the running quark mass ratios, evaluated at $\mu=m_t=171~GeV$, would be as follows
\begin{eqnarray}\label{eq:3.19}
0.000022\leq &\tilde m_u&\leq 0.000037\cr
0.0043\leq &\tilde m_c&\leq 0.0046\cr
0.0013\leq &\tilde m_d&\leq 0.0017\cr
0.032\leq &\tilde m_s&\leq 0.036
\end{eqnarray}
\end{enumerate}
  
The results of the $\chi^2$ fit of the theoretical expressions for $|{V}^{th}_{ij}|$, $J^{th}$, $\alpha^{th}$, $\beta^{th}$ and $\gamma^{th}$ to the experimentally determined quantities is as follows:\\
The quark mixing matrix computed from the theoretical expresion ${\bf V}^{th}$ with the numerical values of quark mass ratios given in (\ref{eq:3.11})and (\ref{eq:3.13}) and the corresponding best values of the symmetry breaking parameter, $Z^{*1/2}=\sqrt{81/32}$, and the CP-violating phase, $\Phi^*=90^{\circ}$, is 
 
\begin{equation}
V^{th}= \pmatrix{
0.9749~e^{i1^{\circ}} & 0.2225~e^{i157^{\circ}} & 0.0036~e^{i85^{\circ}}\cr
0.2223~e^{i113^{\circ}} & 0.9742~e^{i89^{\circ}} & 0.040~e^{i90^{\circ}} \cr
0.0086~e^{i270^{\circ}} & 0.0392~e^{i270^{\circ}} & 0.9992~e^{i90^{\circ}} \cr
}
\label{eq:3.21}
\end{equation}
In order to have an estimation of the sensivity of our numerical results to the uncertainty in the values of the quark mass ratios, we computed the range of values of the matrix of moduli $|V^{th}_{ij}|$, corresponding to the range of values of the mass ratios given in (\ref{eq:3.19}), but keeping $\Phi$ and $Z^{1/2}$ fixed at the values $\Phi^*=90^{\circ}$ and $Z^{*1/2}=\sqrt{81/32}$.
The result is
\begin{equation}
|{\bf V}^{th}|=\pmatrix{
0.9735 - 0.9771 & 0.2151 - 0.2263 & 0.0028 - 0.0040 \cr
0.2151 - 0.2263 & 0.9726 - 0.9764 & 0.037 - 0.043 \cr
0.0078 - 0.0093 & 0.036 - 0.042 & 0.9991 - 0.9993 \cr
},
\label{eq:3.23}
\end{equation}
which is to be compared with the experimentally determined values of the matrix of moduli \cite{ref:3},
\begin{equation}
|{\bf V}^{exp}|=\pmatrix{
0.9742 - 0.9757 & 0.219 - 0.226 & 0.002 - 0.005 \cr
0.219 - 0.225 & 0.9734 - 0.9749 & 0.037 - 0.043 \cr
0.004 - 0.014 & 0.035 - 0.043 & 0.9990 - 0.9993 \cr
}.
\label{eq:3.25}
\end{equation}

As is apparent from (\ref{eq:3.21}), (\ref{eq:3.23}) and (\ref{eq:3.25}), the agreement between computed and experimental values of all entries in the mixing matrix is very good. The estimated range of variation in the computed values of the moduli of the four entries in the upper left corner of the matrix  $|{\bf V}^{th}|$ is larger than the error band in the corresponding entries of the matrix of the experimentally determined values of the moduli $|{\bf V}^{exp}|$.  The estimated range of variation in the computed values of the entries in the third column and the third row of $|{V}^{th}_{ij}|$ is comparable with the error band of the corresponding entries in the matrix of experimentally determined values of the moduli, with the exception of the elements $|V^{th}_{ub}|$ and $|V^{th}_{td}|$ in which case the estimated range of variation due to the uncertainty in the values of the quark mass ratios is significantly smaller than the error band in the experimentally determined value of $|V^{exp}_{ub}|$ and $|V^{exp}_{td}|$.\\

The value obtained for the Jarlskog invariant is 
\begin{equation}\label{eq:3.29}
J^{th}=3.0\times 10^{-5}
\end{equation}
in good agreement with the value $|J^{exp}|=(3.0\pm1.3)\times 10^{-5}sin\delta$ obtained from current data on CP violation in the $K^{\circ}-\bar K^{\circ}$ mixing system \cite{ref:3} and the $B^{\circ}-\bar B^{\circ}$ mixing system \cite{ref:27}.\\

For the inner angles of the unitarity triangle, we found the following central values:
\begin{eqnarray}\label{eq:3.27}
\alpha=83.6^{\circ}\quad\quad\beta=23.2^{\circ}\quad\quad\gamma=73.2^{\circ}.
\end{eqnarray}
An estimation of the range of values of the three inner angles of the unitarity triangle, compatible with the experimental information on the absolute values of the matrix elements ${\bf V}^{exp}$, is given by Mele \cite{ref:32} and Ali \cite{ref:33}. According to this authors, $75^{\circ}\leq\alpha\leq 121^{\circ}$, $16^{\circ}\leq\beta\leq 34^{\circ}$, and $38^{\circ}\leq\gamma\leq 81^{\circ}$. We see that the central value of $\beta$ obtained in this work is close to the lower limit according to Mele \cite{ref:32}, while $\gamma$ is close to the upper limit given by Mele \cite{ref:32} and $\alpha$ is in the allowed range given by these authors.

\section{Phase equivalence of ${\bf V}^{th}$ and the phenomenological parametrizations ${\bf V}^{PDG}$ and ${\bf V}^{KM}$} 
\label{sec:4.0}

In the mass basis, the quark charged currents take the form
\begin{equation}\label{eq:41}
{\it J}^{\mu}_c=\frac{\it g}{\sqrt 2}\bar q^{u}_{Li}\gamma^{\mu}V_{ij}q^{d}_{Lj}.
\end{equation}
A redefinition of the phases of the quark fields which leaves ${\it J}^{\mu}_c$
invariant, will change the arguments of $V_{ij}$ but leave the moduli $|V_{ij}|$ invariant,
\begin{eqnarray}\label{eq:43}
V_{ij}\rightarrow \tilde V_{ij}=e^{-i\chi^u_i}V_{ij}e^{i\chi^d_j}.
\end{eqnarray}

Hence, the invariant measurables of the quark mixing matrix are the moduli of its elements, i.e., the quantities  $|V_{ij}|$, and the Jarlskog invariant J. But even J up to a sign is a function of the moduli \cite{ref:4}. In the case of three families, unitarity of $\bf V$ constrains the number of independent moduli to four \cite{ref:4}. \\
 In consequence, in the current literature, $\bf V$, is usually parametrized in terms of four independent parameters. Kobayashi and Maskawa \cite{ref:1.3} originally chose as independent parameters three rotation angles and one CP-violating phase. A number of parametrizations of this kind, different from the original Kobayashi-Maskawa form have been proposed \cite{ref:1.4}. One of the most commonly used is the ``standard'' parametrization \cite{ref:2} advocated by the Particle Data Book \cite{ref:3}. From a mathematical point of view, all parametrizations of the flavour mixing matrix containing four, suitably defined, independent parameters are equivalent.\\
  In contrast, the parametrization $\bf V^{th}$, derived in the previous sections from the breaking of the flavour permutational symmetry, has only two free, linearly independent parameters, namely, the symmetry breaking parameter $Z^{1/2}$ and the CP-violating phase $\Phi$. When the best values of the parameters   $Z^{1/2}$ and $\Phi$ are used, the mixing matrix $\bf V^{th}$ reproduces the central values of all experimentally determined quantities, that is, the moduli $|V^{exp}_{ij}|$, the Jarlskog invariant $J^{exp}$ and the three inner angles, $\alpha$, $\beta$ and $\gamma$ of the unitarity triangle. The quality of the fit of $\bf V^{th}$ to the experimental data is as good as the quality of the fit of the phenomenological parametrizations  $\bf V^{PDG}$ and  $\bf V^{KM}$ to the same data. More precisely, when the best set of adjustable parameters of each parametrization $\bf V^{th}$, $\bf V^{PDG}$ and  $\bf V^{KM}$, obtained from a $\chi^2$ fit to the same experimental data, the moduli of corresponding entries in the matrices $\bf V^{th}$, $\bf V^{PDG}$ and  $\bf V^{KM}$, are numerically equal and give an equally good representation of the experimentally determined values $|\bf{V^{exp}}_{ij}|$. From this observation, it follows that the symmetry derived $\bf V^{th}$ and the phenomenological parametrizations $\bf V^{PDG}$ and  $\bf V^{KM}$ should be equivalent up to a rephasing of the quark fields in the mass representation. In the following, we will show that it is possible to derive new theoretical mixing matrices $\bf\tilde V^{th}$ and ${\bf\hat{V}}^{th}$, related to $\bf V^{th}$ by biunitary rephasing transformations, and such that all corresponding entries in $\bf\tilde V^{th}$ and $\bf V^{PDG}$ or $\bf{\hat{V}}^{th}$ and $\bf V^{KM}$ are equal in modulus and phase. From here, we will obtain exact, explicit expressions for the three mixing angles and the CP-violating phase ocurring in $\bf V^{PDG}$ and $\bf V^{KM}$ as functions of the quark mass ratios and the parameters  $Z^{*1/2}$ and  $\Phi^*$ characterizing the preferred symmetry breaking pattern.
\section{Phase equivalence of ${\bf V}^{th}$ and ${\bf V}^{PDG}$} 
\label{sec:4}

The standard parametrization ~\cite{ref:2} of the mixing matrix recomended by the Particle Data Group \cite{ref:3} is written in terms of three mixing angles $\theta_{12}, \theta_{23}, \theta_{13}$ and one CP violating phase $\delta_{13}$, 

\begin{equation}\label{eq:4431}
{\bf V}^{PDG}=\pmatrix{
c_{12}c_{13} & s_{12}c_{13} & s_{13}e^{-i\delta_{13}} \cr
-s_{12}c_{23}-c_{12}s_{23}s_{13}e^{i\delta_{13}} & c_{12}c_{23}-s_{12}s_{23}s_{13}e^{i\delta_{13}}& s_{23}c_{13} \cr
s_{12}s_{23}-c_{12}c_{23}s_{13}e^{i\delta_{13}} & -c_{12}s_{23}-s_{12}c_{23}s_{13}e^{i\delta_{13}} & c_{23}c_{13} \cr
}
\end{equation}
where $c_{ij}=\cos\theta_{ij}$ and $s_{ij}=\sin\theta_{ij}$.\\

The range of values of the experimentally determined moduli in $|V^{exp}_{ij}|$, as given by Caso {\it et ~al} \cite{ref:3}, corresponds to $90\%$ confidence limits on the range of values of the mixing angles of

\begin{equation}\label{eq:4433}
0.219\leq s_{12}\leq 0.226,
\end{equation}

\begin{equation}\label{eq:4435}
0.037 \leq s_{23} \leq 0.043,
\end{equation}

\begin{equation}\label{eq:4437}
0.002 \leq s_{13} \leq 0.005.
\end{equation}

Seven of the nine absolute values of the CKM entries have been measured directly, by tree level processes. A range of values for the four parameters, $s_{12}, s_{23}, s_{13}$ and $\delta_{13}$ which is consistent with the seven direct measurements and the experimentally determined values of the moduli of $|{\bf V}|^{exp}$ \cite{ref:3}, is given by Nir \cite{ref:35}

\begin{equation}\label{eq:4438}
0.2173\leq s_{12}\leq 0.2219,
\end{equation}

\begin{equation}\label{eq:4441}
0.0378 \leq s_{23} \leq 0.0412,
\end{equation}

\begin{equation}\label{eq:4443}
0.00237 \leq s_{13} \leq 0.00395,
\end{equation}
$c_{13}$ is known to deviate from unity only in the sixth decimal place [\cite{ref:3}, \cite{ref:35}].\\

 The CP violating phase $\delta_{13}$, at present, is not constrained by direct measurements. However, the measurements of CP violation in $K$ decays \cite{ref:36} force $\delta_{13}$ to lie in the range

\begin{equation}\label{eq:4439}
0 \leq \delta_{13}\leq \pi.
\end{equation}

The standard parametrization ${\bf V}^{PDG}$ was introduced without taking the possible functional relations between the quark masses and the flavour mixing parameters into account. In contrast, these functional relations are explicitly exhibited in the theoretical expressions,  ${V}^{th}_{ij}$, derived in the previous sections. Furthermore, we have seen that, when the best values of the parameters $Z^{1/2}$ and $\Phi$ are used, the mixing matrix ${\bf V}^{th}$ reproduces the central values of all experimentally determined quantities, that is, the moduli $|V^{exp.}_{ij}|$, the Jarlskog invariant $J^{exp.}$ and the three inner  angles, $\alpha$, $\beta$ and $\gamma$, of the unitarity triangle \cite{ref:1}. Since the two parametrizations reproduce the same set of experimental data equally well, we are justified in writing 
\begin{equation}\label{eq:4311}
|V^{th}_{ij}|=|V^{PDG}_{ij}|=|V^{exp}_{ij}|.
\end{equation}
We cannot simply equate ${\bf V}^{th}$ and ${\bf V}^{PDG}$ because the arguments of corresponding matrix elements in the two parametrizations are not equal
\begin{equation}\label{eq:4312}
arg(V^{th}_{ij})\ne arg(V^{PDG}_{ij}).
\end{equation}
This difference is of no physical consequence, it reflects the freedom in choosing the unobservable phases of the quark fields in the mass representation.
 A redefinition of the phases of the quark fields which leaves ${\it J}^{\mu}_c$
invariant, will change the argument of $V^{th}_{ij}$ but leave the moduli $|V^{th}_{ij}|$ invariant,
\begin{eqnarray}\label{eq:44}
V^{th}_{ij}\rightarrow \tilde V^{th}_{ij}=e^{-i\chi^u_i}V^{th}_{ij}e^{i\chi^d_j}.
\end{eqnarray}
The phases $\chi^u_i$ and $\chi^d_j$ ocurring in Eq. (\ref{eq:43}) will be determined from the requirement that corresponding  entries in $\tilde{\bf V}^{th.}$ and ${\bf V}^{PDG}$ be equal,

\begin{equation}\label{eq:45}
 |V^{th}_{ij}|e^{i(w^{th}_{ij}-(\chi^u_i-\chi^d_j))}=|V^{PDG}_{ij}|e^{iw^{PDG}_{ij}},
\end{equation}
in this expression $w^{th}_{ij}$ and $w^{PDG}_{ij}$ are the arguments of $V^{th}_{ij}$ and $V^{PDG}_{ij}$ respectively. Since the moduli $|V^{th}_{ij}|$ and $|V^{PDG}_{ij}|$ are equal, the arguments of the entries in the two parametrizations are related by the set of nine equations 
\begin{equation}\label{eq:47}
\chi^u_i-\chi^d_j=w^{th}_{ij}-w^{PDG}_{ij}.
\end{equation}
The set of Eqs.~(\ref{eq:47}) relate the differences of the unobservable quark field phases to the differences of the arguments of corresponding entries in ${\bf V}^{th}$ and ${\bf V}^{PDG}$. \\
We notice that the set of Eqs. (\ref{eq:47}) is overdetermined. In the left hand side of Eqs. (\ref{eq:47}) there are nine differences of unobsevable quark field phases formed from only six different unknown quark field phases. Since only differences of phases may be determined, the phases themselves are defined only up to an additive constant which may be fixed by giving the value of one of the unknown quark phases. Therefore, in Eqs. (\ref{eq:47}) there are nine equations to determine five unknowns. This is possible only if a set of 4 consistency conditions is satisfied. The consistency conditions are non trivial relations expressing the five non-vanishing arguments $w^{PDG}_{ij}$ of $V^{PDG}_{ij}$  in terms only of the known arguments $w^{th}_{ij}$ of $V^{th}_{ij}$.\\
 A first consistency condition may be derived from the equality of the determinants of $\bf\tilde V^{th}$ and $V^{PDG}$.
From the definition of the rephasing transformation, Eqs.~(\ref{eq:43}) and (\ref{eq:45}), it follows that

\begin{eqnarray}\label{eq:49}
\det{\bf V}^{th}=\det\left[ {\bf X}^\dagger_u {\bf V}^{PDG}{\bf X}_d \right],
\end{eqnarray} 
in this expression ${\bf X}_u$ and ${\bf X}_d$ are the diagonal unitary matrices of phases ocurring in Eq. (\ref{eq:43}). The determinant of ${\bf V}^{PDG}$ is one, hence,
\begin{equation}\label{eq:411}
\det\left[{\bf X}^\dagger_u {\bf V}^{PDG}{\bf X}_d \right]=e^{i\sum^3_{i=1}\left(\chi^{(u)}_i-\chi^{(d)}_i\right)}.
\end{equation}
Similarly, from the definition of ${\bf V}^{th}$, Eq. (\ref{eq:37}), we get
\begin{equation}\label{eq:413}
\det {\bf V}^{th}=\det\left[ {\bf O}^T_u{\bf P}^{u-d}{\bf O}_d\right]=\det\left({\bf O}^T_u{\bf O}_d\right)\det{\bf P}^{u-d},
\end{equation}
the determinant of the orthogonal matrices is one, and the determinant of the diagonal matrix of phases ${\bf P}^{u-d}$ is $e^{i2\Phi}$. Taking for $\Phi$ the best value $\Phi^*=\pi/2$, we obtain

\begin{equation}\label{eq:415}
\det{\bf V}^{th}=e^{i2\Phi^*}=e^{i\pi}.
\end{equation}
 Substitution of Eq. (\ref{eq:411}) and Eq. (\ref{eq:415}) in Eq. (\ref{eq:49}) gives
\begin{equation}\label{eq:417}
\sum^{3}_{i=1}\left( \chi^{(u)}_i-\chi^{(d)}_i\right)=2\Phi^*=\pi.
\end{equation}
This phase relation guarantees the equality of the determinants of $\tilde{\bf V}^{th}$ and ${\bf V}^{PDG}$.\\
The sum of the unobservable quark field phases ocurring in the left hand side of Eq.~(\ref{eq:417}) may be computed from Eqs.~(\ref{eq:47}),
\begin{equation}\label{eq:419}
\sum^{3}_{i=1}\left( \chi^{(u)}_i-\chi^{(d)}_i\right)=\sum^{3}_{i=1}w^{th}_{ii}-w^{PDG}_{22}.
\end{equation}
Now, we eliminate the unobservable quark field phases between Eq.~(\ref{eq:417}) and Eq. (\ref{eq:419}), to get,
\begin{equation}\label{eq:423}
w^{PDG}_{22}=\sum^{3}_{i=1}w^{th}_{ii}-2\Phi^*.
\end{equation}
This,  relation shows that $\arg(V^{PDG}_{22})$ is uniquely determined $(mod ~2\pi)$ in terms of the arguments of the entries in ${\bf V}^{th}$.\\
A set of consistency conditions for the solution of Eqs.~(\ref{eq:47}) may be derived in a similar way by elimination of the quark field phases. From Eqs.~(\ref{eq:47}), differences of phases of the same quark field type, say $\left( \chi^{(d)}_j-\chi^{(d)}_{j'}\right)$, may be computed from Eqs.~(\ref{eq:47}) in at least three different ways. This redundancy implies the existence of non-trivial relations among the arguments of the entries of the two parametrizations. For example, from Eqs.~(\ref{eq:47}), the difference $\left( \chi^{(u)}_2-\chi^{(d)}_3\right)- \left( \chi^{(u)}_2-\chi^{(d)}_2\right)$ gives
\begin{equation}\label{eq:425}
\chi^{(d)}_2-\chi^{(d)}_3=w^{th}_{23}-w^{th}_{22}+w^{PDG}_{22},
\end{equation}
and the difference $\left( \chi^{(u)}_1-\chi^{(d)}_3\right)- \left( \chi^{(u)}_1-\chi^{(d)}_2\right)$ gives
\begin{equation}\label{eq:427}
\chi^{(d)}_2-\chi^{(d)}_3=w^{th}_{13}-w^{th}_{12}+\delta_{13}.
\end{equation}
If the phase difference  $(\chi^{(d)}_2-\chi^{(d)}_3)$ is eliminated between Eqs. (\ref{eq:425}) and (\ref{eq:427}) we get
\begin{equation}\label{eq:429}
\delta_{13}-w^{PDG}_{22}=w^{th}_{12}-w^{th}_{13}-w^{th}_{22}+w^{th}_{23}.
\end{equation}
Using the same elimination procedure for all possible combinations $\left( \chi^{(u)}_i-\chi^{(d)}_{j}\right)-\left( \chi^{(u)}_i-\chi^{(d)}_{j'}\right)$ we derive a set of nine equations, only four of which are linearly independent. 
One of these is Eq. (\ref{eq:429}), for the other three we may take
\begin{equation}\label{eq:431}
-w^{PDG}_{21}+w^{PDG}_{22}=w^{th}_{11}-w^{th}_{12}-w^{th}_{21}+w^{th}_{22},
\end{equation}

\begin{equation}\label{eq:433}
w^{PDG}_{31}-w^{PDG}_{32}=-w^{th}_{11}+w^{th}_{12}+w^{th}_{31}-w^{th}_{32},
\end{equation}
and
\begin{equation}\label{eq:435}
-w^{PDG}_{22}+w^{PDG}_{32}=-w^{th}_{22}+w^{th}_{23}+w^{th}_{32}-w^{th}_{33}.
\end{equation}

Since, in ${\bf V}^{PDG}$ there are five entries with non-vanishing arguments, namely, ~$w^{PDG}_{13}=-\delta_{13},  w^{PDG}_{21}, w^{PDG}_{22}, w^{PDG}_{31}$ and  $w^{PDG}_{32}$, we require still one more equation relating the arguments of the entries of the two parametrizations. This is obtained from the phase relations between the determinants of the two matrices,  ${\bf V}^{th}$  and ${\bf V}^{PDG}$. 
 
With the help of Eq. (\ref{eq:423}) we solve Eqs. (\ref{eq:429})-(\ref{eq:435}) for all the other non-vanishing arguments of ${\bf V}^{PDG}$
\begin{equation}\label{eq:437}
\delta_{13}=w^{th}_{11}+w^{th}_{12}-w^{th}_{13}+w^{th}_{23}+w^{th}_{33}-2\Phi^*
\end{equation}

\begin{equation}\label{eq:439}
w^{PDG}_{21}=w^{th}_{21}+w^{th}_{12}+w^{th}_{33}-2\Phi^*
\end{equation}

\begin{equation}\label{eq:441}
w^{PDG}_{31}=w^{th}_{31}+w^{th}_{12}+w^{th}_{23}-2\Phi^*
\end{equation}

\begin{equation}\label{eq:443}
w^{PDG}_{32}=w^{th}_{32}+w^{th}_{23}+w^{th}_{11}-2\Phi^*.
\end{equation}
In this way, we have shown that the arguments $w^{PDG}_{ij}$ of $V^{PDG}_{ij}$ are uniquely determined (mod ~$2\pi$) by the arguments $w^{th}_{ij}$ of $V^{th}_{ij}$.\\

We now return to the question of the quark field phases and the phase transformation from $V^{th}_{ij}$ to $V^{PDG}_{ij}$.
Substitution of Eqs.~(\ref{eq:423})-(\ref{eq:443}) into Eqs.~(\ref{eq:47}), gives the differences of the quark field phases explicitly in terms of the known arguments $w^{th}_{ij}$ of $V^{th}_{ij}$.   The quark field phases themselves are determined only up to a common additive constant. Since the quark field phases are unobservable, without loss of generality, we may fix one of them, and solve for the others. In this way, if we set $\chi^{d}_{1}=0$, we get
\begin{eqnarray}\label{eq:445}
\chi^{d}_{1}&=&0^\circ,\cr
\chi^{d}_{2}&=&w^{th}_{11}-w^{th}_{12},\cr
\chi^{d}_{3}&=&-w^{th}_{23}-w^{th}_{33}-w^{th}_{12}+2\Phi^*,\cr
\chi^{u}_{1}&=&w^{th}_{11},\cr
\chi^{u}_{2}&=&-w^{th}_{12}-w^{th}_{33}+2\Phi^*,\cr
\chi^{u}_{3}&=&-w^{th}_{23}-w^{th}_{12}+2\Phi^*.
\end{eqnarray}
Then, the diagonal matrices of phases required to compute the phase transformed $\tilde {\bf V}^{th}$ are
\begin{equation}\label{eq:447}
{\bf X}_u=diag[e^{iw^{th}_{11}}, e^{i(-w^{th}_{12}-w^{th}_{33}+2\Phi^*)}, e^{i(-w^{th}_{23}-w^{th}_{12}+2\Phi^*)}]
\end{equation}
and
\begin{equation}\label{eq:449}
{\bf X}_d=diag[1, e^{i(w^{th}_{11}-w^{th}_{12})}, e^{i(w^{th}_{12}-w^{th}_{23}-w^{th}_{33}+2\Phi^*)}].
\end{equation}
Hence, with the help of Eqs. (\ref{eq:437})-(\ref{eq:443}), we verify that
\begin{equation}\label{eq:451}
{\bf X^{\dagger}_u}{\bf V}^{th}{\bf X_d}={\bf V}^{PDG}
\end{equation}
is satisfied as an identity, provided that $|V^{th}_{ij}|=|V^{PDG}_{ij}|$.
\section{The mixing angles of the Standard Parametrization ${\bf V}^{PDG}$ }
\label{sec:5}

The standard parametrization $V^{PDG}_{ij}$ advocated by the PDG and the symmetry derived parametrization $V^{th}_{ij}$, give an equally good representation of the experimentally determined values of the moduli of the entries in the quark mixing matrix $|V^{exp}_{ij}|$ \cite{ref:3}. Hence, we may write
\begin{equation}\label{eq:51}
|V^{th}_{ij}|=|V^{PDG}_{ij}|,
\end{equation}
even though $V^{th}_{ij}$ has only two adjustable parameters $(Z^{1/2},~\Phi)$ while the number of adjustable parameters in $V^{PDG}_{ij}$ is four, namely, $(\theta_{12},~\theta_{23},~\theta_{13},~\delta_{13})$.
All entries in $|V^{th}_{ij}|$ are explicit functions of the four quark mass ratios $(m_{u}/m_{t}, m_{c}/m_{t}, m_{d}/m_{b}, m_{s}/m_{b})$ and the two parameters $Z^{1/2}$ and $\Phi$. The equality of the moduli of corresponding entries of the two parametrizations will allow us to derive explicit expressions for the mixing angles in terms of the four quark mass ratios $(m_{u}/m_{t}, m_{c}/m_{t}, m_{d}/m_{b}, m_{s}/m_{b})$ and the parameters $Z^{1/2}$ and $\Phi$.

From the equality of $|V^{th}_{13}|$ and $|V^{PDG}_{13}|$, it follows that

\begin{equation}\label{eq:53}
\sin\theta_{13}=|V^{th}_{ub}|,
\end{equation}
if we take $|V^{th}_{ub}|$ from (\ref{eq:327}), and we set $\Phi$ and $Z^{1/2}$ equal to their best values $\Phi^*=\pi/2$ and $Z^{1/2*}=\sqrt{\frac{81}{32}}$, we get
\begin{eqnarray}\label{eq:55}
\sin\theta_{13}&=&\bigg\{ \frac{\tilde m_{c}(1-\tilde m_{u}-\delta^{*}_{u})\tilde m_{d}\tilde m_{s}\delta^{*}_{d}}{(1-\delta^*_{u})(1-\tilde m_{u})(\tilde m_{c}+\tilde m_{u})(1-\delta^*_{d})(1+\tilde m_{s})(1-\tilde m_{d})}\cr
&+& \bigg [ \left( \frac{\tilde m_{u}(1-\tilde m_{u}-\delta^{*}_{u})\delta^{*}_{d}}{(1-\tilde m_{u})(\tilde m_{c}+\tilde m_{u})(1+\tilde m_{s})(1-\tilde m_{d})}\right)^{1/2}\cr
&-&\left(\frac{\tilde m_{u}(1+\tilde m_{c}-\delta^{*}_u)\delta^{*}_u(1-\tilde m_{d}-\delta^*_d)(1+\tilde m_s-\delta^*_d)}{(1-\delta^*_u)(1-\tilde m_u)(\tilde m_c +\tilde m_u)(1-\delta^*_d)(1+\tilde m_s)(1-\tilde m_d)}\right)^{1/2}
\bigg ]^{2} \bigg\}^{\frac{1}{2}}
\end{eqnarray}
The computation of $\sin\theta_{23}$ is slightly more involved. From Eq. (\ref{eq:4431}) and the equality of  $|V^{th}_{ij}|$ and $|V^{PDG}_{ij}|$, we obtain

\begin{eqnarray}\label{eq:57}
\sin\theta_{23}=\frac{|V^{PDG}_{cb}|}{\sqrt{1-|V^{PDG}_{ub}|^{2}}}=\frac{|V^{th}_{cb}|}{\sqrt{1-|V^{th}_{ub}|^{2}}}.
\end{eqnarray}
Substitution of the expressions (\ref{eq:331}) and (\ref{eq:327}) with $\Phi^*=\pi/2$ and $Z^{*1/2}=\sqrt\frac{81}{32}$ for $|V^{th}_{cb}|$ and $|V^{th}_{ub}|$ in Eq. (\ref{eq:57}) gives
\begin{eqnarray}\label{eq:59}
\sin\theta_{23}&=&\sqrt{\frac{1-\tilde m_u}{1+\tilde m_c}} 
\bigg\{
\tilde m_u(1+\tilde m_c-\delta^{*}_u)\tilde m_d\tilde m_s\delta^{*}_d+
\bigg [\sqrt{ (1-\delta^*_u)\tilde m_c(1+\tilde m_c-\delta^*_u)(1-\delta^*_d)\delta^*_d}\cr &-&
\sqrt{\tilde m_c(1-\tilde m_u-\delta^*_u)\delta^*_u(1-\tilde m_d-\delta^*_d)(1+\tilde m_s-\delta^*_d)}\bigg ]^2\bigg\}^{1/2}
\cr &\times&
\bigg\{ -\bigg[\sqrt{(1-\delta^*_u)\tilde m_u (1-\tilde m_u-\delta^*_u)(1-\delta^*_d)\delta^*_d}\cr &-&
\sqrt{\tilde m_u(1+\tilde m_c-\delta^*_u)\delta^*_u(1-\tilde m_d-\delta^*_d)(1+\tilde m_s-\delta^*_d)}\bigg]^2 \cr&+&
(1-\delta^*_u)(1-\tilde m_u)(\tilde m_c+\tilde m_u)(1-\delta^*_d)(1+\tilde m_s)(1-\tilde m_d)-\tilde m_c(1-\tilde m_u-\delta^*_u)\tilde m_d\tilde m_s\delta^*_d
\bigg\}^{-1/2}.
\end{eqnarray}
Similarly from Eq. (\ref{eq:4431}) and the equality of $|V^{th}_{12}|$ and $|V^{PDG}_{12}|$, we obtain
\begin{eqnarray}\label{eq:511}
\sin\theta_{12}=\frac{|V^{PDG}_{us}|}{\sqrt{1-|V^{PDG}_{ub}|^{2}}}=\frac{|V^{th}_{us}|}{\sqrt{1-|V^{th}_{ub}|^{2}}}.
\end{eqnarray}
Then, substitution of the expressions (\ref{eq:325}) and (\ref{eq:327}) for $|V^{th}_{us}|$ and $|V^{th}_{ub}|$ in (\ref{eq:511}) gives
\begin{eqnarray}\label{eq:513}
\sin\theta_{12}&=&\sqrt{\frac{1-\tilde m_d}{\tilde m_s+\tilde m_d}} 
\bigg\{
\tilde m_c(1-\tilde m_u-\delta^{*}_u) \tilde m_d(1+\tilde m_s-\delta^{*}_d)
\cr &+&
\bigg [\sqrt{ (1-\delta^*_u)\tilde m_u(1-\tilde m_u-\delta^*_u)(1-\delta^*_d)\tilde m_s(1+\tilde m_s-\delta^*_d)}\cr &+&\sqrt{\tilde m_u(1+\tilde m_c-\delta^*_u)\delta^*_u\tilde m_s(1-\tilde m_d-\delta^*_d)\delta^*_d}\bigg ]^2\bigg\}^{1/2}\cr &\times&
\bigg\{-\bigg[\sqrt{(1-\delta^*_u)\tilde m_u (1-\tilde m_u-\delta^*_u)(1-\delta^*_d)\delta^*_d}\cr&-&
\sqrt{\tilde m_u(1+\tilde m_c-\delta^*_u)\delta^*_u(1-\tilde m_d-\delta^*_d)(1+\tilde m_s-\delta^*_d)}\bigg]^2\cr&+&
(1-\delta^*_u)(1-\tilde m_u)(\tilde m_c+\tilde m_u)(1-\delta^*_d)(1+\tilde m_s)(1-\tilde m_d)\cr &-&
\tilde m_c(1-\tilde m_u-\delta^*_u)\tilde m_d\tilde m_s\delta^*_d \bigg\}^{-1/2}.
\end{eqnarray}
The computed values for $\sin\theta_{12},\sin\theta_{23}$ and $\sin\theta_{13}$ corresponding to the best $\chi^2$ fit of $|V^{th}_{ij}|$, $J^{th}$ and $\alpha^{th}$, $\beta^{th}$ and $\gamma^{th}$ to the experimentally determined quantities $|V^{exp}_{ij}|$, $J^{exp}$ and the three inner angles of the unitarity triangle $\alpha^{exp}$, $\beta^{exp}$ and $\gamma^{exp}$ are obtained when the numerical values of $|V^{th}_{us}|$, $|V^{th}_{ub}|$ and $|V^{th}_{cb}|$ computed from Eqs.(\ref{eq:325}), (\ref{eq:327}), (\ref{eq:331}) and given in Eq.~(\ref{eq:3.21}) are substituted in to Eqs.~(\ref{eq:53}), (\ref{eq:57}) and (\ref{eq:511}). In this way, we get
\begin{eqnarray}\label{eq:515}
\sin\theta^*_{12}=0.222,
\end{eqnarray}
\begin{eqnarray}\label{eq:517}
\sin\theta^*_{23}=0.040,
\end{eqnarray}
\begin{eqnarray}\label{eq:519}
\sin\theta^*_{13}=0.0036.
\end{eqnarray}
The numerical value of $cos\theta^*_{13}$ deviates from unity in the sixth decimal place.\\

We notice that the numerical values of the mixing angles computed from quark masses and the best values of the symmetry breaking parameters coincide almost exactly with the central values of the experimentally determined quantities, as could be expected from Eq.~(\ref{eq:51}). This observation is interesting because, in the case of three families, the most general form of the mixing matrix has at most four free, independent parameters \cite{ref:4} which could be four independent moduli or three mixing angles and one phase as occurs in  ${\bf V}^{PDG}$. The symmetry derived ${\bf V}^{th}$ has only two free, real independent parameters. In spite of that, the quality of the fit of ${\bf V}^{th}$ to the experimental data is as good as the quality of the fit of ${\bf V}^{PDG}$ to the same data. The predictive power of ${\bf V}^{th}$ implied by this fact originates in the flavour permutational symmetry of the Standard Model and the assumed symmetry breaking pattern  from which the texture in the quark mass matrices and ${\bf V}^{th}$ were derived. 
\section{The Cp violating phase $\delta_{13}$}
\label{sec:6}
The CP violating phase $\delta_{13}$ of the standard parametrization ${\bf V}^{PDG}$ of the quark mixing matrix is given in Eq. (\ref{eq:437}) in terms of the arguments $w^{th}_{ij}$ of five entries in the theoretical expression for  $V^{th}_{ij}$ and the corresponding CP violating phase $\Phi$. Taking from Eq. (\ref{eq:3.21}) the numerical values of the arguments  $w^{th}_{ij}$ and setting $\Phi$ equal to the best value $\Phi^*=\pi/2$, we obtain the numerical value of $\delta_{13}$ corresponding to the best fit of $|V^{th}_{ij}|$ to the experimental data
\begin{equation}\label{eq:61}
\delta^*_{13}=73.2^{\circ}.
\end{equation}
This predicted value of  $\delta_{13}$ is very close to the numerical value of the third inner angle $\gamma$, of the unitarity triangle. The difference may readily be computed in terms of the arguments  $w^{th}_{ij}$. From the expression for $\gamma$ 
\begin{equation}\label{eq:62}
\gamma=arg\left[ -\frac{V^{*}_{cb}V_{cd}}{V^{*}_{ub}V_{ud}}\right]
\end{equation}
we get
\begin{equation}\label{eq:63}
-\gamma=w^{th}_{11}-w^{th}_{13}-w^{th}_{21}+w^{th}_{23}+\pi
\end{equation}
which, when compared with the expression (\ref{eq:437}) for $\delta_{13}$ gives
\begin{equation}\label{eq:65}
-\gamma=\delta_{13}-(w^{th}_{12}+w^{th}_{21}+w^{th}_{33}-2\Phi^*-\pi).
\end{equation}
Taking from Eq. (\ref{eq:3.21}) the numerical values of the arguments corresponding to the best values $\Phi^*=90^{\circ}$ and $Z^{1/2}*= \sqrt{\frac{81}{32}}$, we obtain
\begin{equation}\label{eq:67}
(w^{th}_{12}+w^{th}_{21}+w^{th}_{33}-2\Phi^*-\pi)=0.04^{\circ}.
\end{equation}
This is, indeed, a very small number, and justifies the approximation
\begin{equation}\label{eq:69}
-\gamma \approx \delta^*_{13}.
\end{equation}
According to this, the value of $|\gamma|$ computed from quark mass ratios and the best values of the parameters $Z^{*1/2}$ and $\Phi^*$ is $|\gamma|=73.2^{\circ}$, in agreement with the bounds extracted from the precise measurements of the $B^0_d$ oscillation frequency \cite{ref:32} and the measurements of the rates of the exclusive hadronic decays $B^{\pm}\rightarrow \pi K$ and the CP averaged $B^{\pm}\rightarrow \pi^{\pm}\pi^0$ \cite{ref:37}.
Exact explicit expressions for the CP violating phase $\delta_{13}$ in terms of the four quark mass ratios and the parameters $Z^{*1/2}$ and $\Phi^*$ may readily be found; such an expression could be derived from Eq. (\ref{eq:437}) in terms of the arguments of five matrix elements of ${\bf V}^{th}$. However, a simpler expression, involving only four matrix elements of ${\bf V}^{th}$ may be obtained from the Jarlskog invariant $J$. \\

   The Jarlskog invariant may be written in terms of four matrix elements of ${\bf V}$ as
\begin{equation}\label{eq:611}
J={\it Im}[V_{12}V_{23}V^*_{13}V^*_{22}].
\end{equation}
Since $J$ is an invariant, its value is independent of the particular parametrization of ${\bf V}$. If we write the right hand side of Eq. (\ref{eq:611}) in terms of the standard parametrization ${\bf V}^{PDG}$, we obtain

\begin{equation}\label{eq:613}
\sin\delta_{13}=\frac{J^{th}}{s_{12}s_{13}s_{23}c_{12}c^2_{13}c_{23}}.
\end{equation} 
The terms in the denominator in the right hand side of this expression were written in Eqs. (\ref{eq:53}), (\ref{eq:57}) and (\ref{eq:511}) in terms of the moduli $|V^{th}_{12}|$, $|V^{th}_{13}|$ and $|V^{th}_{23}|$. Hence,
\begin{equation}\label{eq:615}
s_{12}s_{13}s_{23}c_{12}c^2_{13}c_{23}=\frac{|V^{th}_{12}||V^{th}_{13}||V^{th}_{23}|[(1-|V^{th}_{13}|^2-|V^{th}_{12}|^2)(1-|V^{th}_{13}|^2-|V^{th}_{23}|^2)]^{1/2}}{1-|V^{th}_{13}|^2}.
\end{equation}
Substitution of Eq. (\ref{eq:615})  in Eq. (\ref{eq:613}) gives
\begin{equation}\label{eq:617}
\sin{\delta_{13}}=\frac{(1-|V^{th}_{13}|^2){\it Im}[V^{th}_{12}V^{th}_{23}V^{th*}_{13}V^{th*}_{22}]}
{|V^{th}_{12}||V^{th}_{13}||V^{th}_{23}|\sqrt{ (1-|V^{th}_{13}|^2-|V^{th}_{12}|^2)(1-|V^{th}_{13}|^2-|V^{th}_{23}|^2)}},
\end{equation}
the right hand side of this equation may be written in terms of the quark mass ratios and the symmetry breaking parameters $Z^{*1/2}$ and $\Phi^*$ with the help of Eqs.  (\ref{eq:55}), (\ref{eq:59}) and (\ref{eq:513}).\\
A simpler expression which leads to a very accurate approximation for $\delta_{13}$ is obtained from Eq. (\ref{eq:617}) if the matrix elements in the square brackets are written as modulus and argument, and use is made of the unitarity of ${\bf V}^{th}$ to simplify the denominator,
\begin{equation}\label{eq:619}
\sin{\delta_{13}}=\frac{(1-|V^{th}_{13}|^2)|V^{th}_{22}| \sin{(w^{th}_{12}+w^{th}_{23}-w^{th}_{13}-w^{th}_{22})}}
{ |V^{th}_{11}||V^{th}_{33}|}.
\end{equation}
Explicit expressions for the arguments $w^{th}_{12}, w^{th}_{23}, w^{th}_{13}$ and $w^{th}_{22}$ in terms of the quark mass ratios may be derived from  Eqs. (\ref{eq:325})-(\ref{eq:331}) setting $Z^{1/2}$ and $\Phi$ equal to their best values $Z^{*1/2}=\sqrt{\frac{81}{32}}$ and $\Phi^*=\pi/2$, we get
\begin{equation}\label{eq:621}
w^{th}_{us}=\pi-\tan^{-1}\left( \sqrt{ \frac{\tilde m_u\tilde m_s}{\tilde m_c\tilde m_d}}\left[\sqrt{(1-\delta^*_u)(1-\delta^*_d)}+\sqrt{ \delta^*_u\delta^*_d\frac{(1+\tilde m_c-\delta^*_u)(1-\tilde m_d-\delta^*_d)}{(1-\tilde m_u-\delta^*_u)(1+\tilde m_s-\delta^*_d)}}\right]\right),
\end{equation}

\begin{equation}\label{eq:623}
w^{th}_{cb}=\pi-\tan^{-1}\left( \sqrt{ \frac{\tilde m_c}{\tilde m_u\tilde m_d\tilde m_s}}\left[\sqrt{(1-\delta^*_u)(1-\delta^*_d)}-\sqrt{\frac{\delta^*_u(1-\tilde m_u-\delta^*_u)(1-\tilde m_d-\delta^*_d)(1+\tilde m_s-\delta^*_d)}{\delta^*_d(1+\tilde m_c-\delta^*_u)}}\right]\right),
\end{equation}

\begin{equation}\label{eq:625}
w^{th}_{ub}=\tan^{-1}\left( \sqrt{ \frac{\tilde m_u}{\tilde m_c\tilde m_d\tilde m_s}}\left[\sqrt{(1-\delta^*_u)(1-\delta^*_d)}-\sqrt{\frac{\delta^*_u(1+\tilde m_c-\delta^*_u)(1-\tilde m_d-\delta^*_d)(1+\tilde m_s-\delta^*_d)}{\delta^*_d(1-\tilde m_u-\delta^*_u)}}\right]\right),
\end{equation}

\begin{equation}\label{eq:627}
w^{th}_{cs}=\tan^{-1}\left( \sqrt{ \frac{\tilde m_c\tilde m_s}{\tilde m_u\tilde m_d}}\left[\sqrt{(1-\delta^*_u)(1-\delta^*_d)}+\sqrt{ \delta^*_u\delta^*_d\frac{(1-\tilde m_u-\delta^*_u)(1-\tilde m_d-\delta^*_d)}{(1+\tilde m_c-\delta^*_u)(1+\tilde m_s-\delta^*_d)}}\right]\right).
\end{equation}

Computing the second factor in square brackets in the leading order of magnitude, we get

\begin{equation}\label{eq:629}
w^{th}_{us}\approx\pi-\tan^{-1}\left( \sqrt{ \frac{\tilde m_u\tilde m_s}{\tilde m_c\tilde m_d}}\right),
\end{equation}

\begin{equation}\label{eq:631}
w^{th}_{cb}\approx\pi-\tan^{-1}\left( \sqrt{ \frac{\tilde m_c}{\tilde m_u\tilde m_d\tilde m_s}}\left[(1-\sqrt{\frac{\delta^*_u}{\delta^*_d}})\right]\right),
\end{equation}

\begin{equation}\label{eq:633}
w^{th.}_{ub}\approx\tan^{-1}\left( \sqrt{ \frac{\tilde m_u}{\tilde m_c\tilde m_d\tilde m_s}}\left[(1-\sqrt{\frac{\delta^*_u}{\delta^*_d}})\right]\right),
\end{equation}
and
\begin{equation}\label{eq:635}
w^{th}_{cs}\approx\tan^{-1}\left( \sqrt{ \frac{\tilde m_c\tilde m_s}{\tilde m_u\tilde m_d}}\right).
\end{equation}
 The modulus $|V^{th}_{ub}|$ has already been expressed in terms of quark mass ratios and the parameters characterizing the symmetry breaking pattern $Z^{*1/2}$ and $\Phi^*$, in Eqs. (\ref{eq:53}) and (\ref{eq:55}). Similar expressions for the other moduli ocurring in Eq. (\ref{eq:619}) may also be given

\begin{eqnarray}\label{eq:637}
|V_{ud}|&=&\left(\frac{\tilde {m}_c \left( 1-\tilde{m}_u -\delta^*_u \right)
\tilde {m}_s \left( 1-\tilde{m}_d -\delta^*_d \right)}{\left( 1-\delta^*_u \right)\left( 1-\tilde{m}_u \right)\left( \tilde {m}_c+\tilde {m}_u \right)
\left( 1-\delta^*_d \right)\left( 1-\tilde{m}_d \right)
\left( \tilde {m}_s+\tilde {m}_d \right)}\right)^{1/2}\cr &\times&\left\{1+
\frac{\tilde{m}_u \tilde{m}_d}{\tilde{m}_c \tilde{m}_s}\left[\left((1-\delta^*_u)(1-\delta^*_d) \right)^{1/2}+\left( \delta^*_u\delta^*_d\frac{(1+\tilde {m}_c-\delta^*_u)(1+\tilde {m}_s-\delta^*_d)}{(1-\tilde {m}_u-\delta^*_u)(1-\tilde {m}_d-\delta^*_d)}\right)^{1/2}\right]^2\right\},
\end{eqnarray}

\begin{eqnarray}\label{eq:639}
|V_{cs}|&=&\left(\frac{\tilde {m}_c
\left( 1+\tilde{m}_c -\delta^*_u \right)\tilde {m}_s
\left( 1+\tilde{m}_s -\delta^*_d \right)}{
\left( 1+\tilde{m}_c \right)\left( \tilde {m}_c+\tilde {m}_u \right)
\left( 1+\tilde{m}_s \right)
\left( \tilde {m}_s+\tilde {m}_d \right)}\right)^{1/2}\cr &\times&
\bigg\{\left[1+
\left(\frac{\delta^*_u\delta^*_d\left( 1-\tilde {m}_u-\delta^*_u \right)
\left( 1-\tilde{m}_d -\delta^*_d \right)}{\left( 1-\delta^*_u \right)\left( 1-\delta^*_d \right)
\left( 1+\tilde{m}_c-\tilde{m}_u \right)
\left( 1+\tilde{m}_s-\delta^*_d \right)}\right)^{1/2}\right]^2 \cr&+&
\frac{\tilde{m}_u \tilde{m}_d}{\tilde{m}_c \tilde{m}_s}\frac{1}{( 1-\delta^*_u )( 1-\delta^*_d)}
\bigg\}^{1/2},
\end{eqnarray}

\begin{eqnarray}\label{eq:641}
|V_{tb}|&=&\left[\frac{(1-\tilde{m}_u -\delta^*_u)(1+\tilde{m}_c -\delta^*_u)
(1-\tilde{m}_d -\delta^*_d)(1-\tilde m_s -\delta^*_d) }
{(1- \delta^*_u)(1+\tilde{m}_c)(1-\tilde{m}_u)
(1- \delta^*_d )(1+\tilde{m}_s)(1-\tilde{m}_d)}
\right]^{1/2} 
\cr &\times&\bigg\{\left[1+\left(\frac{\delta^*_u\delta^*_d(1-\delta^*_u)(1- \delta^*_d )}
{(1+\tilde{m}_c-\delta^*_u)(1-\tilde{m}_u-\delta^*_u)
(1-\tilde{m}_d-\delta^*_d)(1+\tilde{m}_s-\delta^*_d)}\right)^{1/2}\right]^2\cr
&+& \frac{\tilde {m}_u \tilde {m}_c \delta^*_u 
\tilde {m}_d \tilde{m}_s \delta^*_d  }
{(1-\tilde {m}_u- \delta^*_u )(1 + \tilde {m}_c- \delta^*_u )(1+ \tilde {m}_s- \delta^*_d)(1-\tilde{m}_d-\delta^*)}\bigg\}^{1/2}.
\end{eqnarray}

Computing in the leading order of magnitude, the first factor in the right hand side of Eq. (\ref{eq:619}) gives
\begin{equation}\label{eq:643}
\frac{(1-|V^{th}_{13}|^2)|V^{th}_{22}| }{ |V^{th}_{11}||V^{th}_{33}|}\approx \frac{(1-\delta^*_u)(1-\tilde m_u)(1-\delta^*_d)(1-\tilde m_d)}{(1-\tilde m_u-\delta^*_u)(1-\tilde m_d-\delta^*_d)}\left( 1-\frac{\tilde m_u}{\tilde m_d}(\sqrt{\delta^*_d}-\sqrt{\delta^*_u})\right).
\end{equation}
Inserting in to Eq.~(\ref{eq:643}) the numerical values of the mass ratios and $\sqrt{\delta^*_d}-\sqrt{\delta^*_u}=0.04$, we find that the right hand side of Eq.(\ref{eq:643}) differs from one in the third decimal place, 
\begin{equation}\label{eq:645}
\frac{(1-|V^{th}_{13}|^2)|V^{th}_{22}| }{ |V^{th}_{11}||V^{th}_{33}|}\approx 1.
\end{equation}
Therefore, 
\begin{equation}\label{eq:647}
\sin\delta^*_{13}\approx \sin(w^{th}_{us}+w^{th}_{cb}-w^{th}_{ub}-w^{th}_{cs}),
\end{equation}
taking the numerical values of the argument in the right hand side of Eq. (\ref{eq:647}) from (\ref{eq:3.21}), we obtain
\begin{equation}\label{eq:649}
\delta^*_{13}\approx 73^{\circ},
\end{equation}
in agreement with Eq. (\ref{eq:61}).
The approximate expression Eq. (\ref{eq:647}) for $\sin\delta^*_{13}$ could also be derived from Eq. (\ref{eq:429}) if $w^{PDG}_{22}$ is neglected. Computing $w^{PDG}_{22}$ from Eq. (\ref{eq:423}) and (\ref{eq:621})-(\ref{eq:627}), we obtain $w^{PDG}_{22}=-0.0018^{\circ}$ which shows that Eq. (\ref{eq:647}) is a very good approximation. Since Eq. (\ref{eq:423}) was derived from the the phase relations expressing the arguments of $V^{PDG}_{ij}$  in terms of those of  $V^{th}_{ij}$, while Eq. (\ref{eq:619})  was derived from the expression Eq. (\ref{eq:617}) for the Jarlskog invariant, the agreement found between  Eqs. (\ref{eq:423}) and  (\ref{eq:619})- (\ref{eq:647}) is a consistency check of our formalism.

\section{Phase equivalence of ${\bf V}^{th}$ and the Kobayashi-Maskawa parametrization ${\bf V}^{KM}$}
\label{sec:7}

The quark mixing matrix was parametrized by Kobayashi and Maskawa \cite{ref:1.3} in terms of the three mixing angles, $\theta_{1}$, $\theta_{2}$, and $\theta_{3}$, and one CP violating phase $\delta_{13}$

\begin{equation}\label{eq:771}
{\bf V}^{KM}=\pmatrix{
c_{1} & -s_{1}c_{3} & -s_{1}s_{3} \cr
s_{1}c_{2} & c_{1}c_{2}c_{3}-s_{2}s_{3}e^{i\delta_{KM}}& c_{1}c_{2}s_3+s_2 s_3 e^{i\delta_{KM}} \cr
s_{1}s_{2}& c_{1}s_{2}c_3-c_{2}s_{3}e^{i\delta_{KM}} & c_{1}s_{2}s_{3-}-c_{2}s_{3}e^{i\delta_{KM}} \cr
}
\end{equation}
where $c_{i}=\cos\theta_{i}$ and $s_{i}=\sin\theta_{i}$.\\

It is readily verified that 

\begin{equation}\label{eq:773}
\det{\bf V}^{KM}=-e^{i\delta_{KM}}
\end{equation}

  As discussed in section \ref{sec:4.0}, the parametrization ${\bf V}^{th}$ derived from the breaking of the flavour symmetry and the Kobayashi-Maskawa parametrization ${\bf V}^{KM}$, give an equally good representation of the values of the moduli $|{\bf V}_{ij}^{exp}|$ of the mixing matrix even if ${\bf V}^{th}$ has only two free, linearly independent parameters. Hence, the two parametrizations are equivalent up to a rephasing of the quark fields. Therefore, we may define a phase transformed ${\hat{\bf V}}^{th}$, such that all entries in  ${\hat {\bf V}}_{ij}^{th}$ are numerically equal to the corresponding entries in  ${\bf V}_{ij}^{KM}$ both in modulus and phase,

\begin{equation}\label{eq:775}
{\hat{\bf V}}^{th}={\bf X}_{KM}^{u\dagger}{\bf V}^{th}{\bf X}_{KM}^{d}={\bf V}^{KM}.
\end{equation}

The diagonal matrices

\begin{equation}\label{eq:777}
{\bf X}_{KM}^{u}=diag[e^{i\phi_1^u}, e^{i\phi_2^u} e^{i\phi_3^u}] 
\end{equation}
 
and 

\begin{equation}\label{eq:779}
{\bf X}_{KM}^{d}=diag[e^{i\phi_1^d}, e^{i\phi_2^d} e^{i\phi_3^d}] 
\end{equation}

are determined from the equality of ${\hat{\bf V}}^{th}$ and ${\bf V}_{ij}^{KM}$,

\begin{equation}\label{eq:7711}
\phi_i^u-\phi_j^d=w_{ij}^{th}-w_{ij}^{KM}.
\end{equation}

where

\begin{equation}\label{eq:7713}
w_{ij}^{KM}=arg({\bf V}_{ij}^{KM}).
\end{equation}

  In the left hand side of Eqs.~(\ref{eq:7711}) there are nine differences of phases formed from only six unobservables quark field phases. Differences of phases of the same quark field type, say $(\phi_j^d - \phi_{j'}^d)$, may be computed from Eqs. (\ref{eq:7711}) in at least three different ways. Elimination of the unobservables differences of quark field phases between these expressions gives a set of consistency conditions relating the known $w_{ij}^{th}$ and $w_{ij}^{KM}$. 

\begin{equation}\label{eq:7715}
w_{12}^{KM}=\pi,
\end{equation}
\begin{equation}\label{eq:7717}
w_{13}^{KM}=\pi,
\end{equation}

\begin{equation}\label{eq:7719}
w_{22}^{KM}=w_{11}^{th}-w_{12}^{th}-w_{21}^{th}+w_{22}^{th}+\pi,
\end{equation}

\begin{equation}\label{eq:7721}
w_{23}^{KM}=w_{11}^{th}-w_{13}^{th}-w_{21}^{th}+w_{23}^{th}+\pi,
\end{equation}

\begin{equation}\label{eq:7723}
w_{32}^{KM}=w_{11}^{th}-w_{12}^{th}-w_{31}^{th}+w_{32}^{th}+\pi,
\end{equation}

\begin{equation}\label{eq:7725}
w_{33}^{KM}=w_{11}^{th}-w_{13}^{th}-w_{31}^{th}+w_{33}^{th}+\pi,
\end{equation}
 when the expressions (\ref{eq:7715}-\ref{eq:7725}) are substituted in Eqs. (\ref{eq:7711}) we obtain the differences of the quark field phases as functions of the arguments $w_{ij}^{th}$ of ${\bf V}_{ij}^{th}$. The quark field phases themselves are determined only up to a common additive constant which is fixed by giving the value of one of them. If we set $\chi_i^u=0$, we get

\begin{equation}\label{eq:7727}
{\bf X_{KM}^u}=diag[1, e^{-i(w_{11}^{th}-w_{21}^{th})}, e^{-i(w_{11}^{th}-w_{31}^{th})}]
\end{equation}

and 

\begin{equation}\label{eq:7729}
{\bf X_{KM}^d}=diag[e^{-iw_{11}^{th}}, e^{-i(w_{12}^{th}-\pi)}, e^{-i(w_{13}^{th}-\pi)}].
\end{equation}

With the help of (\ref{eq:7727}) and (\ref{eq:7729}) it is readily verified 
that  (\ref{eq:775}) is satisfied as an identity.

\section{The CP violating phase $\delta_{KM}$ of the Kobayashi-Maskawa parametrization.}
\label{sec:7.1}

The CP violating phase $\delta_{KM}$ is implicitly determined by the equations (\ref{eq:7715}~-~\ref{eq:7725}). An explicit expression for $\delta_{KM}$ in terms of the arguments $w_{ij}^{th}$ of ${\bf V}_{ij}^{th}$ may be obtained from the identity

\begin{equation}\label{eq:7731}
\det [{\bf V}^{KM}]=\det [{\hat{\bf V}}^{th}],
\end{equation}

and

\begin{equation}\label{eq:7733}
\det{[{\hat{\bf V}}^{th}]}=e^{-i\sum_{i=1}^3 (\phi_i^u-\phi_i^d)}e^{i2\Phi^*}.
\end{equation}
Substitution of (\ref{eq:773}) and (\ref{eq:7733}) in (\ref{eq:7731}) gives

\begin{equation}\label{eq:7735}
\delta_{KM}=\sum_{i=1}^3(\phi_i^d-\phi_i^u)+2\Phi^*-\pi
\end{equation}

 Now, from Eqs. (\ref{eq:7711}) and (\ref{eq:7715}~-~\ref{eq:7725}), 

\begin{equation}\label{eq:7737}
\delta_{KM}=w_{ud}^{th}-w_{us}^{th}-w_{ub}^{th}-w_{cd}^{th}-w_{td}^{th}.
\end{equation}

  This expression gives the CP-violating phase $\delta_{KM}$ of the Kobayashi-Maskawa parametrization,  ${\bf V}^{KM}$, in terms of the known arguments of the symmetry derived parametrization ${\bf V}^{th}$.\\
Taking the numerical values of the arguments $w_{ij}^{th}$ ocurring in (\ref{eq:7737}) from Eq. (\ref{eq:3.21}), we obtain 

\begin{equation}\label{eq:7739}
\delta_{KM}=96.4^{\circ}
\end{equation}

or
\begin{equation}\label{eq:7741}
\delta_{KM}=\pi - 83.6^{\circ}.
\end{equation}

The value of the inner angle $\alpha$ of the unitarity triangle found in our $\chi^2$ fit of ${\bf V}^{th}$ to the experimentally determined values of the moduli $|{\bf V}_{ij}^{exp}|$ is

\begin{equation}\label{eq:7742}
\alpha=83.6^{\circ}.
\end{equation}

 Therefore, within the acuracy of our best fit to the experimental data 
 
\begin{equation}\label{eq:7743}
\delta_{KM}\approx \pi - \alpha.
\end{equation}

This is only an approximation, to derive an exact relation we compute $\alpha$ from 

\begin{equation}\label{eq:7745}
\alpha= \arg\left( -\frac{V_{ub}^*V_{ud}}{V_{tb}^*V_{td}}\right),
\end{equation}

substitution of ${\bf V}_{ij}^{th}$ for ${\bf V}_{ij}$ in Eq. (\ref{eq:7745}) gives

\begin{equation}\label{eq:7747}
\alpha=\pi-w_{ub}^{th}+w_{ud}^{th}+w_{tb}^{th}-w_{td}^{th}.
\end{equation}

Comparing (\ref{eq:7747}) with (\ref{eq:7737}) gives

\begin{equation}\label{eq:7749}
\delta_{KM}=-\alpha-(w_{us}^{th}+w_{cd}^{th}-w_{tb}^{th}+\pi).
\end{equation}

Taking the numerical values of the $w_{ij}^{th}$ from (\ref{eq:3.21}) we get

\begin{equation}\label{eq:7751}
w_{us}^{th}+w_{cd}^{th}+w_{tb}^{th}+\pi=3\pi+0.04^{\circ}.
\end{equation}

Hence,

\begin{eqnarray}\label{eq:7753}
\delta_{KM}=\pi-\alpha-0.04^{\circ}\quad\quad mod(2\pi)
\end{eqnarray}

 which shows that (\ref{eq:7743}) is an excellent approximation to the numerical value of  $\delta_{KM}$.\\
 In passing, we notice that

\begin{equation}\label{eq:7755}
\alpha=w_{tb}^{KM}
\end{equation}
is an exact result.

\section{The mixing angles in the parametrization of Kobayashi-Maskawa.}
\label{sec:7.2}

Once it is established that when the best set of parameters of each parametrization, ${\bf V}^{KM}(\theta_1, ~\theta_2, ~\theta_3, ~\delta_{KM})$ and ${\bf V}^{th}(\tilde m_i, ~Z^{1/2}, ~\Phi)$ is used, the moduli of corresponding entries in the two parametrizations are numerically equal, we may identify corresponding entries and solve for the mixing angles $\theta_1, ~\theta_2$ and  $\theta_3$ in terms  of the quark mass ratios and the parameters $\Phi$ and $Z$.\\

  From the identification

\begin{equation}\label{eq:7757}
\cos\theta_1=|{\bf V}_{ud}^{th}|,
\end{equation}

we obtain

\begin{eqnarray}\label{eq:7759}
\cos\theta_1&=&\left(\frac{(1-\tilde m_u-\delta_u^*)(1-\tilde m_d-\delta_d^*)}{(1-\delta_u^*)(1-\tilde m_u)(1-\tilde m_d)(1-\delta_d^*)(1+\tilde m_u/\tilde m_c)(1+\tilde m_d/\tilde m_s)}\right)^{1/2}\cr &\times&
\left[ 1+\frac{\tilde m_u\tilde m_d}{\tilde m_c \tilde m_s}\left( \left( (1-\delta_u^*)(1-\delta_d^*)\right)^{1/2}+\left( \delta_u^*\delta_d^* \frac{ (1+\tilde m_c-\delta_u^*)(1+\tilde m_s-\delta_d^*)}{ (1-\tilde m_u-\delta_u^*)(1+\tilde m_d-\delta_d^*)} \right)^{1/2}\right)^2\right]^{1/2}
\end{eqnarray}

The angle $\theta_2$ is obtained from the identification

\begin{equation}\label{eq:7765}
\sin\theta_1 \sin\theta_2=|{\bf V}_{td}^{th}|
\end{equation}

which gives

\begin{equation}\label{eq:7767}
\sin\theta_2=\frac{|{\bf V}_{td}^{th}|}{\sin\theta_1}
\end{equation}

where
\begin{eqnarray}\label{eq:7766}
\sin\theta_1=\sqrt{1-|{\bf V}_{ud}^{th}|^2},
\end{eqnarray}
$|{\bf V}_{ud}^{th}|$ is given in Eqs. ~(\ref{eq:7757}) and (\ref{eq:7759}), 
and
\begin{eqnarray}\label{eq:7768}
{\bf V}_{td}^{th}&=&\left( \frac{\tilde {m}_u\tilde m_c \delta_u\tilde m_s
(1-\tilde{m}_d -\delta_d ) }{ (1-\delta_u)
(1+\tilde{m}_c)(1-\tilde{m}_u)
(1-\delta_d)(1 - \tilde {m}_d)(\tilde m_s +\tilde m_d) } \right)^{1/2}
\cr
&+&\bigg\{- \left(\frac{(1-\tilde{m}_u -\delta_u)(1+\tilde{m}_c -\delta_u)
\tilde{m}_d (1+\tilde{m}_s -\delta_u)\delta_d }
{(1- \delta_u)(1+\tilde{m}_c)(1-\tilde{m}_u)
(1- \delta_d )(1-\tilde{m}_d)(\tilde{m}_s +\tilde{m}_d)}
\right)^{1/2}
\cr &+& \left(\frac{\delta_u\tilde{m}_d (1-\tilde{m}_d -\delta_d)}
{(1+\tilde{m}_c)(1-\tilde{m}_u)(1-\tilde{m}_d)(\tilde{m}_s +\tilde{m}_d)}
\right)^{1/2}\bigg\}e^{i\Phi}  
\end{eqnarray}

Computing in the leading order, we obtain
\begin{eqnarray}\label{eq:7769}
\sin\theta_2 &\approx& \sqrt{\frac{1+m_u/m_c}{1+\frac{m_um_s}{m_cm_d}}}Z^{1/2}
\bigg\{\frac{\tilde m_s-\tilde m_d}{\sqrt{(1-\tilde m_d)(1+\tilde m_s)+2Z(\tilde m_s -\tilde m_d)(1+\frac{(\tilde m_s -\tilde m_d)}{2})}}\cr &-&
\frac{\tilde m_c-\tilde m_u}{\sqrt{(1-\tilde m_u)(1+\tilde m_c)+2Z(\tilde m_c -\tilde m_u)(1+\frac{(\tilde m_c -\tilde m_u)}{2})}}\bigg\}
\end{eqnarray}

similarly the mixing angle $\theta_3$ may be obtained from the identification
\begin{equation}\label{eq:7775}
\sin\theta_1\sin\theta_3=|{\bf V}_{ub}^{th}|
\end{equation}
Then,
\begin{equation}\label{eq:7777}
\sin\theta_3=\frac{|{\bf V}_{ub}^{th}|}{\sqrt{1-|{\bf V}_{ud}^{th}|^2}}
\end{equation}
where ${\bf V}_{ub}^{th}$ is given in Eq. (\ref{eq:327}) and $\sin\theta_1$ is given in Eq. (\ref{eq:7766}).
Computing in the leading order of magnitude, we find,
\begin{eqnarray}\label{eq:7779}
\sin\theta_3 &\approx& \sqrt{\frac{1+m_d/m_s}{1+\frac{m_dm_c}{m_sm_u}}}Z^{1/2}
\bigg\{\frac{\tilde m_s-\tilde m_d}{\sqrt{(1-\tilde m_d)(1+\tilde m_s)+2Z(\tilde m_s -\tilde m_d)(1+\frac{(\tilde m_s -\tilde m_d)}{2})}}\cr &-&
\frac{\tilde m_c-\tilde m_u}{\sqrt{(1-\tilde m_u)(1+\tilde m_c)+2Z(\tilde m_c -\tilde m_u)(1+\frac{(\tilde m_c -\tilde m_u)}{2})}}\bigg\}
\end{eqnarray}

The exact values computed for $\sin{\theta_1}$, $\sin{\theta_2}$ and $\sin{\theta_3}$ corresponding to the best fit of ${\bf V}_{ij}^{th}$, $J^{th}$ and $\alpha^{th}$, $\beta^{th}$ and $\gamma^{th}$ to the experimentally determined quantities are obtained from Eqs. (\ref{eq:7757}, ~\ref{eq:7767}) and (\ref{eq:7777}) when the numerical values of ${\bf V}_{ud}^{th}$, ${\bf V}_{td}^{th}$ and ${\bf V}_{ub}^{th}$ computed from Eqs. (\ref{eq:637}), (\ref{eq:7759}) and (\ref{eq:7768}) are substituted. We obtain
\begin{equation}\label{eq:7785}
\sin{\theta_1}=0.2225
\end{equation}
\begin{equation}\label{eq:7787}
\sin{\theta_2}=0.0384
\end{equation}
\begin{equation}\label{eq:7789}
\sin{\theta_3}=0.0162
\end{equation}
and, from Eq. (\ref{eq:7739}),

\begin{equation}\label{eq:7791}
\sin{\delta_{KM}}=0.9938.
\end{equation}

If the mixing angles $\theta_1$$, \theta_2$ and $\theta_3$ are restricted to lie in the first quadrant, the corresponding numerical values of the angles are

\begin{eqnarray}\label{eq:7792}
\theta_1=12.86^{\circ}\quad\theta_2=2.2^{\circ}\quad and\quad\theta_3=0.93^{\circ}
\end{eqnarray}

which together with the numerical value of the CP-violating phase $\delta_{KM}$ found in the previous section, $\delta_{KM}=96.4^{\circ}$,  gives the best set of values of mixing parameters in the Kobayashi-Maskawa parametrization of the mixing matrix corresponding to the best set of parameters $\Phi^*=90^{\circ}$  and $Z^{*1/2}=\sqrt{81/32}$ of the flavour symmetry breaking derived parametrization ${\bf V}^{th}$ of the mixing matrix.\\
As expected, from the way they were obtained, the numerical values of the moduli $|{\bf V}_{ij}^{KM}|$ computed with the help of these numerical values of the mixing angles $\theta_1$, $\theta_2$,  $\theta_1$, and $\delta_{KM}$ reproduce the central values  of the experimentally determined $|{\bf V}_{ij}^{exp}|$, given in Caso  $\it et~al$. \cite{ref:3}\\
In the Kobayashi-Maskawa parametrization, the Jarlskog invariant is 

\begin{equation}\label{eq:7793}
J=s_1^2s_2s_3c_1c_2c_3s_{\delta}
\end{equation}
The corresponding numerical value is
\begin{equation}\label{eq:7795}
J=3.0\times 10^{-5}
\end{equation}
in excellent agrement with $J^{exp}$.

\section{Summary and Conclusions}
\label{sec:11}
In the Standard Electroweak model of particle interactions, both, the masses of the quarks as well as the mixing parameters and the CP-violating phase appear as free independent parameters. In consequence, phenomenological parametrizations of the quark mixing matrix were introduced without taking the possible functional relations between the quark masses and the flavour mixing parameters into account. These functional relations are explicitly exhibited in the theoretical quark mixing matrix ${\bf V}^{th}$ derived from the breaking of the flavour permutational symmetry in previous works \cite{ref:1}, \cite{ref:1.1} and reviewed in sections \ref{sec:2} and \ref{sec:3} of this paper. \\

In this work, we explicitly exhibit the phase equivalence of the theoretical mixing matrix, ${\bf V}^{th}$, and two phenomenological parametrizations: the original Kobayashi-Maskawa \cite{ref:1.3}, ${\bf V}^{KM}$, and the standard parametrization  ${\bf V}^{PDG}$ advocated by the Particle Data Group \cite{ref:2}, \cite{ref:3}, which are written in terms of three mixing angles and one CP-violating phase.\\
The equality of the moduli of corresponding entries in ${\bf V}^{th}$  and  ${\bf V}^{KM}$ or ${\bf V}^{PDG}$ allowed us to derive, exact, explicit expressions for the mixing angles and the CP-violating phase of the two phenomenological parametrizations as functions of four quark mass ratios $m_u/m_t$, $m_c/m_t$, $m_d/m_b$, $m_s/m_b$, and the flavour symmetry breaking parameters: $Z^{1/2}$ and $\Phi$.\\

The numerical values of the mixing parameters of the PDG advocated standard parametrization $\sin\theta_{12}^*, ~\sin\theta_{23}^*, ~\sin\theta_{13}^*$ computed from the quark mass ratios and the best values of the parameters $Z^{*1/2}$ and $\Phi^*$, coincide almost exactly with the central values of the same mixing parameters determined from a fit to the experimental data \cite{ref:35}, as could be expected from the phase equivalence of ${\bf V}^{th}$ and ${\bf V}^{PDG}$, and the good agreement of $|{\bf V}_{ij}^{th}|$ with the corresponding experimentally determined values  $|{\bf V}_{ij}^{exp}|$.\\
  Similarly, from the equality of the moduli of corresponding entries in ${\bf V}^{th}$ and ${\bf V}^{KM}$, we derived exact, explicit expressions for the Kobayashi-Maskawa mixing parameters, $\sin\theta_{1}, ~\sin\theta_{2}$ and $\sin\theta_{3}$ as functions of the four quark mass ratios and the flavour symmetry breaking parameters $Z^{1/2}$ and $\Phi$. As in the previous case, the numerical values of the mixing parameters, $\sin\theta_{1}^*, ~\sin\theta_{2}^*$ and $\sin\theta_{3}^*$ and the CP-violating phase computed from our expressions and the quark mass ratios and the best values,  $Z^{*1/2}$ and $\Phi^*$, of the symmetry breaking parameters, are such that the numerical values of $|{\bf V}_{ij}^{KM}|$ reproduce almost exactly the central values of the experimentally determined values of the moduli, $|{\bf V}_{ij}^{exp}|$, and the Jarlskog invariant, $J^{exp}$, as given in Caso ${\it et. ~al.}$ \cite{ref:3}.\\
The numerical values of the CP-violating phase $\delta_{13}=arg\left({\bf V}_{ub}^{*PDG}\right)$ computed from quark mass ratios and the best values of the parameters $Z^{*1/2}$ and $\Phi^*$ is  $\delta_{13}=73.2^{\circ}$ in good agreement with the bounds extracted from the precise measurements of the $B_d^o$ oscillations  frequency \cite{ref:32} and the measurements of the rates of the exclusive hadronic decays $B^{\pm}\rightarrow \pi K$ \cite{ref:37}. The numerical values of the CP-violating phase $\delta_{KM}$ of the Kobayashi-Maskawa parametrization computed from quark mass ratios and the best values of $Z^{1/2}$ and $\Phi$ is $\delta_{KM}^* =96.4^{\circ}$. It may be worth to remark that  $\delta_{KM}$ is not a small number as it is sometimes assumed \cite{ref:38}-\cite{ref:39}.\\

In conclusion, the three mixing angles and the CP-violating phase which appear in the phenomenological parametrization of the quark mixing matrix as free, linearly independent parameters are expressed in this work as functions of four quark mass ratios and two flavours symmetry breaking parameters $Z^{1/2}$ and $\Phi$. The numerical values of the mixing angles and CP-violating phase computed from quark mass ratios and the best values of the symmetry breaking parameters $Z^{*1/2}=\sqrt{81/32}$ and $\Phi^*=90^{\circ}$ coincide almost exactly with the central values of the same mixing parameters determined from a fit to the experimental data. The predictive power of the theoretical quark mixing matrix ${\bf V}^{th}$ implied by this fact has its origin in the flavour permutational symmetry of the Standard Model and the assumed symmetry breaking pattern from which the texture in the quark mass matrices and the quark mixing matrix ${\bf V}^{th}$ were derived.

\section*{Acknowledgments}
We are indebted to Prof. M. D. Scadron for some useful discussions and critical remarks.
This work was partially supported by DGAPA-UNAM under contract No. PAPIIT-IN125298 and by CONACYT (M\'exico) under contract 32238E.

\end{document}